\documentclass[twocolumn,aps,prb,superscriptaddress,floatfix, nofootinbib]{revtex4-2}
\usepackage[T1]{fontenc}
\usepackage[utf8]{inputenc}
\usepackage[english]{babel}
\usepackage{color}
\usepackage{float}
\usepackage{braket}
\usepackage{amsmath}
\usepackage{blindtext}
\usepackage{amstext}
\usepackage{tikz}
\usetikzlibrary{patterns}
\usepackage{appendix}
\usepackage{amssymb}
\usepackage{graphicx}
\usepackage{enumerate}
\usepackage{bbold}
\usepackage{bm}
\usepackage{dsfont}
\usepackage[unicode=true,pdfusetitle,
bookmarks=true,bookmarksnumbered=false,bookmarksopen=false,
breaklinks=false,pdfborder={0 0 1},backref=false,colorlinks=false]
{hyperref}
\usepackage{ulem}

\hypersetup{
	colorlinks,linkcolor=blue,citecolor=blue,urlcolor=blue}
\usepackage{soul}

\newcommand{\ppa}[1]{\textcolor{red}{\fbox{Patrícia} {\sl#1}}}

\newcommand{\jc}[1]{\textcolor{magenta}{\fbox{João} {\sl#1}}}

\newcommand{\tc}[1]{\textcolor{blue}{\fbox{Tarik} {\sl#1}}}

\begin{document}

\preprint{APS/123-QED}

\title{Anisotropic resonance energy transfer with strained phosphorene}

\author{J. Oliveira-Cony}
\email{joaooctavio8@gmail.com}
\author{C. Farina}%
\affiliation{Universidade Federal do Rio de Janeiro, Rio de Janeiro, RJ}
\author{P. P. Abrantes}
\affiliation{Universidade Federal do Rio de Janeiro, Rio de Janeiro, RJ}
\author{Tarik P. Cysne}
\email{tarik.cysne@gmail.com}
\affiliation{Universidade Federal Fluminense, Niterói, RJ}

\date{\today}

\begin{abstract}
We analyze the resonance energy transfer (RET) rate between quantum emitters (QEs) near a phosphorene/SiC interface under the effects of uniaxial strain. Using a low-energy tight-binding model, we describe the electronic structure of strained phosphorene in an experimentally feasible situation. Due to the anisotropic electronic structure of phosphorene, we demonstrate that the RET rate drastically depends on the direction in which the QEs are separated relative to the phosphorene lattice. More specifically, we obtain a large variation in the RET rate when the QEs are separated along the zigzag direction, in contrast to a rather small variation when separated along the armchair direction of phosphorene's crystalline structure. Furthermore, our results reveal that the RET rate can be highly modulated by uniaxial strain in phosphorene when considering emitters placed along the zigzag direction. Finally, by means of a simple toy model, we also show that this anisotropy in the RET rate is a general characteristic produced by anisotropic 2D materials. 
\end{abstract}

\maketitle


\section{\label{Introduction}Introduction}

Quantum electrodynamics, which describes radiation-matter interactions, is one of the most precise theories in physics, as the agreement between theoretical predictions and observed experimental data has achieved an unprecedented level \cite{Morgner2023, Aguillard2024, Loetzsch2024}. It can be used to describe processes not only in high-energy particle physics but also in low-energy scales, such as when non-relativistic atoms interact with the radiation field in its vacuum state. These latter processes are usually referred to as quantum vacuum effects~\cite{milonni2013}. 
 
The omnipresence of the vacuum electromagnetic field gives rise to various interesting and unexpected phenomena \cite{milton2001casimir, dalvit2011, Novotny_Hecht_2012, dodonov2020}, including the so-called resonance energy transfer (RET) \cite{Andres1987, Jones2019, Andrews2021}. The simplest system in which RET typically occurs is constituted by two closely spaced atoms, one excited and the other in the ground state, where the radiation field has no excitations (i.e., no photons). Since this initial state is not an eigenstate of the total Hamiltonian, it will inevitably evolve in time, leading to a nonvanishing probability of finding another configuration in which the excitation energy from atom $\mathcal{A}$ will be transferred to atom $\mathcal{B}$ without photon emission.

RET is a phenomenon historically related to biology and chemistry. The discovery of RET by the German physical chemist Theodor F\"orster \cite{Forster1946} was in the context of light-sensitive molecules while studying the unexpectedly high efficiency of photosynthesis. More recently, the ability to harness RET has become a sought-after goal due to its wide range of applications, such as the engineering of photo and biosensors \cite{ Chen2012, Geißler2013, Hussain2014, Yang2016, Verma2023, Ha2024}, spectroscopic nano-rulers \cite{Sahoo2011, Baryshnikova2015,Zhang2023}, supramolecular systems \cite{Teunissen2018, Mayoral2018, Raydev2019}, luminescent solar concentrators \cite{Banal2017, TUMMELTSHAMMER2017, Zhang2022}, among many others \cite{Gorbenko2017, Bartnik2019, Fan2023, Li2024}.

Over the last few decades, great theoretical advances and technological developments in nanoplasmonics and nanophotonics have allowed astonishing control of radiation-matter interactions at increasingly smaller scales. This feat can be achieved by changing the neighborhood of the quantum emitters (QEs), which alters the boundary conditions of the vacuum electromagnetic field modes \cite{Yablonovitch1994, KortKamp2014, KortKamp2015, Reinaldo2015, Szilard2019, chinh2020, Casabone2021, Agarwal2024}. Particularly, this strategy has been extensively used for controlling RET, as shown in Refs.~\cite{Barnes1998, Barnes1999, Jones2019, Abrantes2020, patricia2021, Nayem2023, Lezhennikova2023, Beutler2024}.

In this sense, 2D materials stand out from conventional ones due to their highly flexible electronic properties, which reveal unique light-matter coupling behaviors \cite{Plantey2021, Meng2023}. Numerous nanophotonic effects in 2D materials are being studied theoretically and experimentally, including the physics of dispersive forces \cite{Banishev2013, Cysne2014, Woods2016, RodriguezLopez2017,  Marcus2019, Liu2021, Abrantes2021b}, near-field effects in QEs \cite{Gaudreau2013, Raja2016, patricia2021, Nayem2023, Patricia2024, Cavicchi2024}, and heat transfer \cite{Zhao2017, Ge2019, Wu2019, Iqbal2023}, to mention a few. The ever-growing class of these materials holds significant promise for future practical applications.

Phosphorene, a monolayer of black phosphorus first synthesized in 2014 \cite{Li2014, Liu2014}, has since become a particularly attractive 2D material for nanophotonic applications \cite{Lu2016}, as its electronic band structure presents a direct bandgap \cite{Rudenko2014, Menezes2018} that can be controlled by uniform strain \cite{ Rodin2014, Taghizadeh2016, Midtvedt2017}. Furthermore, phosphorene exhibits anisotropic conductivity properties that enable rich physics, including the presence of hyperbolic plasmon modes \cite{Nemilentsau2016, Veen2019}. In fact, many studies have unveiled the potential applications of phosphorene in nanophotonics \cite{Sun2017, Thiyam2018, Mu2021, Sikder2022, Nayem2023, Tao2024, Cavicchi2024}, including the tunability of spontaneous emission of electric and magnetic QEs by strain engineering~\cite{Patricia2024}. As a matter of fact, this tool may provide an easier control parameter in experiments compared to others, such as external fields. 

In this work, we benefit from strain as an alternative method for modulating RET. By manipulating the optomechanical properties of strained phosphorene, we show that the RET rate between two QEs near phosphorene can be modified in a feasible experimental setup. We also highlight how the anisotropy of the optical tensors of phosphorene plays a remarkable role in this system. For QEs placed along the zigzag direction, the RET rate becomes much more sensible than along the armchair direction. Lastly, using a simple toy model, we also show that similar behavior is expected in a generic anisotropic media. 


\section{\label{RETFormalism} Mathematical Formalism}

\begin{figure}[b]
    \centering    \includegraphics[width=0.8\linewidth]{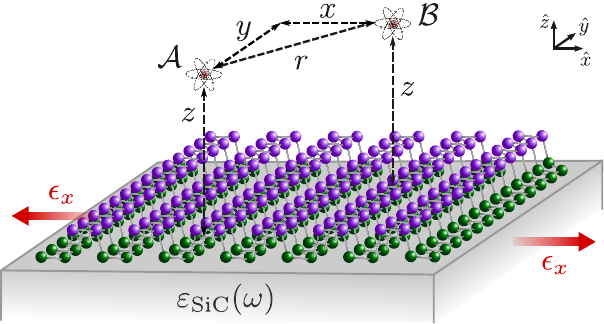}
    \caption{Two QEs separated by a distance $r$ and both at a distance $z$ from a phosphorene/SiC interface, in which uniaxial strain may be applied. In this sketch, strain $\epsilon_x$ is applied along the $x-$direction (armchair).}
    \label{Boneco}
\end{figure}

The system under study is depicted in Fig.~\ref{Boneco}. The donor QE $\mathcal{A}$ is in the excited state $\ket{e}$ and the acceptor QE $\mathcal{B}$ is in the ground state $\ket{g}$, so that the initial configuration can be represented by the vector state $\ket{\mathcal{A}, \mathcal{B}; \mathcal{F}_{{\bm k}p}} = \ket{e,g;0_{{\bm k}p}}$, where $\ket{\mathcal{F}_{{\bm k}p}} = \ket{0_{{\bm k}p}}$ denotes the absence of photons in the electromagnetic field. After evolving in time, the final state of the system becomes $\ket{g,e;0_{{\bm k}p}}$.  In our system, the QEs are located at positions ${\bm r}_{A} = (x_A, y_A, z_A)$ and ${\bm r}_{B} = (x_B, y_B, z_B)$ ($x_{B} - x_{A} =: x$ and $y_{B} - y_{A} =: y$) above a half-space region formed by a phosphorene sheet grown on top of a substrate made of silicon carbide (SiC). The presence of the substrate allows the application of strain to the phosphorene, illustrated along the $x-$direction in Fig.~\ref{Boneco}.

The optical properties of this medium are encoded in the Fresnel reflection matrix
\begin{eqnarray}
    \mathds{R}=\sum_{\alpha,\beta\in\{s,p\}}r^{\beta,\alpha}\boldsymbol{\epsilon}_\alpha^{+}\otimes \boldsymbol{\epsilon}_\beta^{-}, \label{RMatrix}
\end{eqnarray}
where the indices $s$ and $p$ correspond to the TE and TM modes of the electromagnetic field, respectively \cite{Born2019}. We define the unitary vectors
\begin{eqnarray}
&&\boldsymbol{\epsilon}_s^{+}=\boldsymbol{\epsilon}_s^{-}=\dfrac{-k_y\hat{x}+k_x\hat{y}}{k_{\parallel}},\hspace{2mm} \\
&&\boldsymbol{\epsilon}_p^{\pm}=\dfrac{\pm ck_z(k_x\hat{x}+k_y\hat{y})+ck_{\parallel}^2\hat{z}}{k_{\parallel}\omega} \, {,}
\end{eqnarray}
where $\omega=2\pi c/\lambda$ is the angular frequency of {the} electromagnetic mode with wavelength $\lambda$ and ${\bm k}_{\parallel}=k_x\hat{x}+k_y\hat{y}$. 

To calculate the RET rate, we use the dyadic Green function formalism by solving the Helmholtz equation
\begin{eqnarray}
[(\boldsymbol{\nabla}_{{\bm r}} \times\boldsymbol{\nabla}_{{\bm r}}\times) - k^2({\bm r},\omega)]\mathds{G}({\bm r},{\bm r}';\omega)= \delta({\bm r}-{\bm r}')\mathds{1}
\end{eqnarray}

\noindent for the planar geometry described above \cite{ Novotny_Hecht_2012,Amorim2017}, where $k^2(\omega, {\bm r})=\epsilon(\omega,{\bm r})\omega^2/c^2$ and $\epsilon(\omega,\bm r)$ is the dimensionless electrical permittivity. Considering only electric dipole transitions and modeling the emitters within the electric dipole approximation -- i.e., the field obeys ${\bm E}({\bm r};\omega)=\mathds{G}({\bm r},{\bm r}';\omega)\cdot {\bm d}({\bm r}';\omega)$ --, the normalized RET rate is 
\begin{equation}
    \frac{\Gamma_{\rm RET}}{\Gamma_{\mathrm{RET}}^{(0)}} = \frac{\big|{\bm d}_B^{eg}\cdot \mathds{G}({\bm r}_B, {\bm r}_A;\omega)\cdot {\bm d}_A^{ge} \big|^2}{\big|{\bm d}_B^{eg}\cdot \mathds{G}^{(0)}({\bm r}_B;{\bm r}_A;\omega)\cdot {\bm d}_A^{ge} \big|^2}\,, \label{RETgeneral}
\end{equation}
where $\Gamma_{\mathrm{RET}}^{(0)}$ stands for the RET rate in free space, $\omega$ is the QEs transition frequency and ${\bm d}^{ij}_{A,B} = \bra{i,j; 0_{{\bm k}p}} \hat{\bm d}_{A,B}\ket{j,i;0_{{\bm k}p}}$ are matrix elements of the electric dipole operator that describe the atomic transition.

One may express the dyadic Green function $\mathds{G}({\bm r}_B, {\bm r}_A;\omega)$ as a sum of its free space contribution $\mathds{G}^{(0)}({\bm r}_B, {\bm r}_A;\omega)$ and its scattered part $\mathds{G}^{(S)}({\bm r}_B,{\bm r}_A;\omega)$, such that
\begin{equation}
    \mathds{G}(\textbf{r}_B,\textbf{r}_A;\omega)=\mathds{G}^{(0)}(\textbf{r}_B,\textbf{r}_A;\omega)+\mathds{G}^{(S)}(\textbf{r}_B,\textbf{r}_A;\omega)\,.
\end{equation}
The free space Green function is obtained from \cite{Novotny_Hecht_2012}
\begin{equation}
    \mathds{G}^{(0)}({\bm r}, {\bm r}';\omega)=\left[\mathds{1}+\frac{1}{k({\bm r}, \omega)}\nabla_{\bm r}\nabla_{{\bm r}'}\right]G_0({\bm r},{\bm r}';\omega)\,, \label{G0}
\end{equation}
where $G_0({\bm r},{\bm r}';\omega)=\left( 4\pi|{\bm r}-{\bm r}'|\right)^{-1}\mathrm{e}^{ik({\bm r}, \omega)|{\bm r}-{\bm r}'|}$, while the scattered Green function can be written as
\begin{equation}
    \mathds{G}^{(S)}({\bm r}_B,{\bm r}_A;\omega)=\frac{i}{2}\int\frac{\,d^2{\bm k}_{\parallel}}{(2\pi)^2}\mathds{R}\frac{\mathrm{e}^{i[{\bm k}_{\parallel}\cdot({\bm r}_B-{\bm r}_A)+k_z(z_A+z_B)]}}{k_z}\,. \label{GS}
\end{equation}

Throughout this paper, we assume that both transition dipole moments are oriented along the $z-$axis, thereby yielding 
\begin{eqnarray}
    \mathds{G}_{zz}^{(0)}({\bm r}_B,{\bm r}_A;\omega)&=\dfrac{\mathrm{e}^{i\omega r/c}}{4\pi r}\left[1-\left(\dfrac{c}{\omega r}\right)^2+\dfrac{ic}{\omega r}\right] \,,
\end{eqnarray}    
where $r := |\bm{r}_B - {\bm r}_A|$, and
\begin{eqnarray}
&&\mathds{G}_{zz}^{(S)}({\bm r}_B,{\bm r}_A;\omega)=\frac{ic^2}{4\pi }\int d^2\textbf{k}_{\parallel}\,k_{\parallel}^2 r^{p,p}  \nonumber \\
&& \hspace{25mm} \times\frac{\mathrm{e}^{i[\textbf{k}_{\parallel}\cdot(\textbf{r}_B-\textbf{r}_A)+k_z(z_B+z_A)]}}{\omega^2k_z}\,.
\end{eqnarray}

\noindent In this situation, the normalized RET rate can be cast into the form
\begin{equation}
    \dfrac{\Gamma_{\rm RET}}{\Gamma_{\rm RET}^{(0)}} = \left|1+\dfrac{\mathds{G}_{zz}^{(S)}(\textbf{r}_B,\textbf{r}_A;\omega)}{\mathds{G}_{zz}^{(0)}(\textbf{r}_B,\textbf{r}_A;\omega)}\right|^2. \label{RETzz}
\end{equation}
\noindent Generalization to other dipole orientations is straightforward and follows analogously.

\section{\label{Control} Controlling RET with Strained Phosphorene}

The conductivity tensor of strained phosphorene reads
\begin{eqnarray}
\overleftrightarrow{\bm{\sigma}} (\omega, \epsilon_{\mu})=\begin{bmatrix}
\sigma_{xx}(\omega, \epsilon_{\mu}) & 0  \\
0 & \sigma_{yy}(\omega, \epsilon_{\mu})\
\end{bmatrix} \,,
\label{TensorSigma}
\end{eqnarray}

\noindent where $\epsilon_{\mu}$ ($\mu = x, y, z$) is the uniform strain applied along the $\mu-$direction, and the longitudinal conductivities along the $x$ and $y$ axes are computed using linear response theory (see Appendix \ref{ConductivitiesSec} for more details). 


\subsection{RET on relaxed phosphorene\label{subsecA}}

\begin{figure}[b]
    \centering
    \includegraphics[width=1\linewidth]{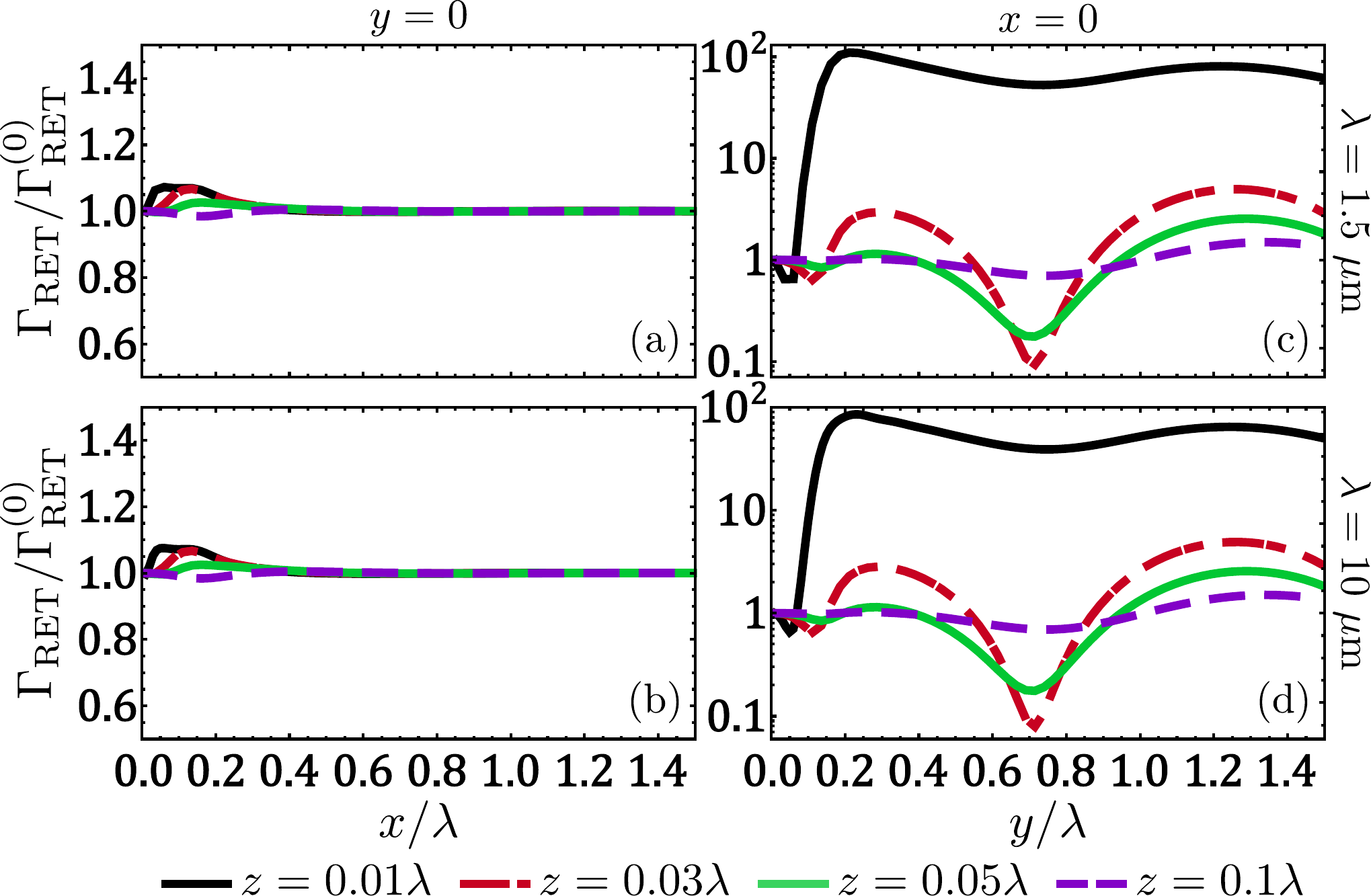}
    \caption{Normalized RET rate as a function of the distance between the QEs placed along the $x-$axis [panels (a) and (b)] and $y-$axis [panels (c) and (d)]. We set $\lambda=1.5$~$\mu$m in the first row and $\lambda=10$~$\mu$m in the second row.}
    \label{phosphoreneNoStrainXY}
\end{figure}

For now, we will be concerned with how phosphorene influences the RET rate without applying strain. In addition, to explore the impacts that the anisotropy of phosphorene may have on the RET rate, we analyze different configurations of the system: {\it (i)} separating the QEs along the armchair direction, i.e., maintaining $y=0$ and varying $x$ [Figs.~\ref{phosphoreneNoStrainXY}(a) and~\ref{phosphoreneNoStrainXY}(b)], and {\it (ii)} separating the QEs along the zigzag direction, i.e., maintaining $x=0$ and varying $y$ [Figs.~\ref{phosphoreneNoStrainXY}(c) and~\ref{phosphoreneNoStrainXY}(d). This analysis is carried out for different distances $z$ between the emitters and phosphorene and for two transition wavelengths ($\lambda=1.5,10$ $\mu$m). The Fermi energy is set at $E_{\rm F}=0.7$~eV and we used the energy scales associated with the electronic scattering process $\eta_1=\eta_2=25$~meV \cite{Patricia2024} in all the results presented in this paper. 

We call the reader's attention to three points. {\it (i)} All plots obey the limit that the normalized RET rate must converge to $1$ as $x$ and $y$ tend to $0$. The closer the QEs are to each other, the less relevant the effects of phosphorene on the energy transfer process, as expected. {\it (ii)}  All curves have an oscillation pattern, commonly present in previous works dealing with RET near non-ideal surfaces \cite{Karanikolas2016, Ding2017, Wu2018, Lezhennikova2023}.  
{\it (iii)}  While the normalized RET rate barely changes for QEs separated along the $x-$axis (armchair), it is drastically changed when the separation is along the $y-$axis (zigzag).

\begin{figure}[t]
    \centering
    \includegraphics[width=1\linewidth]{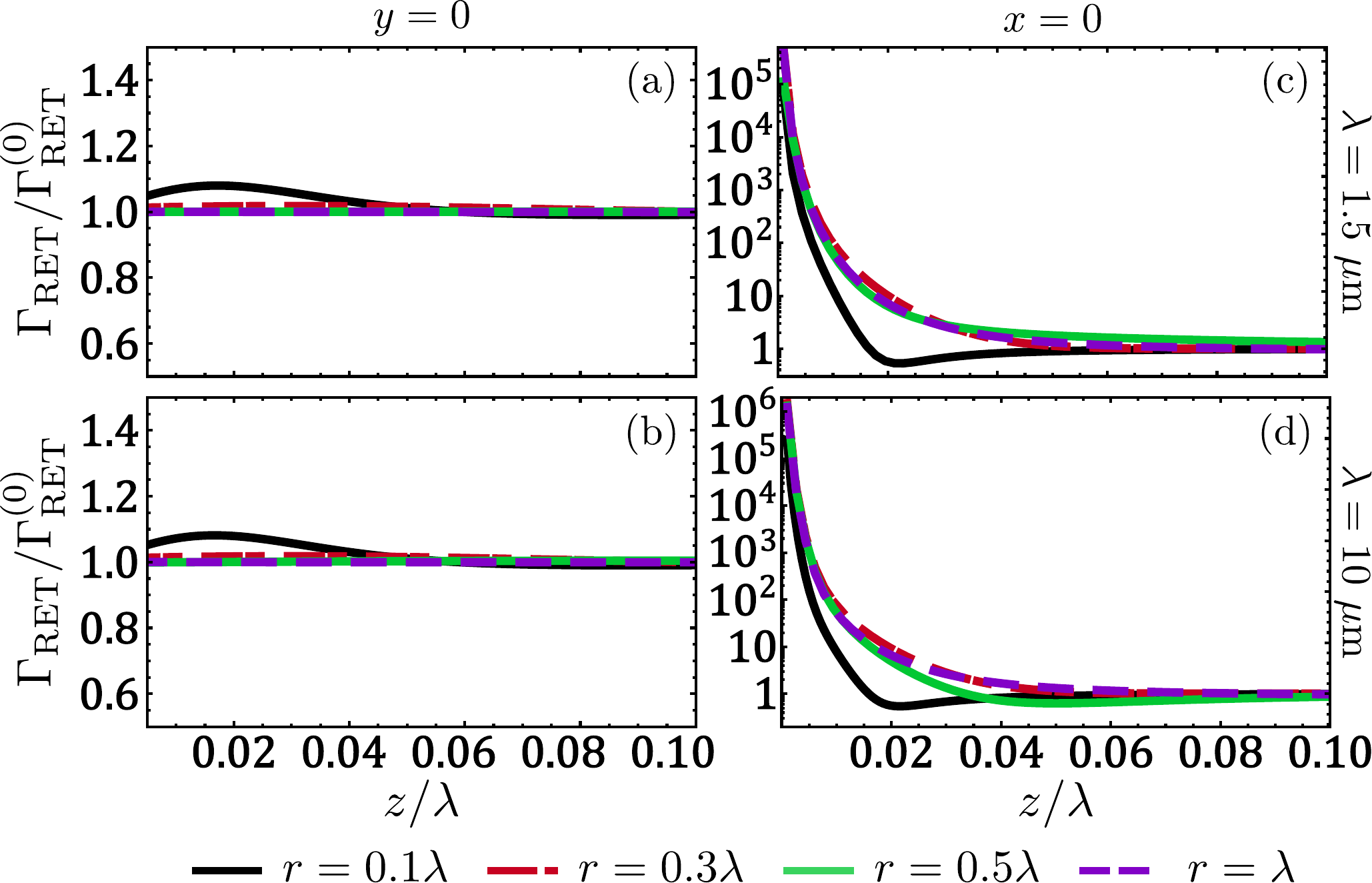}
    \caption{Normalized RET rate as a function of $z$ for different distances between the QEs placed along the $x-$axis [panels (a) and (b)] and $y-$axis [panels (c) and (d)]. We set $\lambda=1.5$~$\mu$m in the first row and $\lambda=10$~$\mu$m in the second row.}
    \label{phosphoreneNoStrainZ}
\end{figure}

In Fig.~\ref{phosphoreneNoStrainZ}, we do the complementary plots, fixing the distance $x$ or $y$ between both QEs and varying the distance between them to the phosphorene sheet. In this case, the analog of comment {\it (i)} is that the normalized RET rate must converge to $1$ when $z \rightarrow \infty$, as the influence of phosphorene is progressively reduced. These results also underline how the position of the QEs to the phosphorene is crucial and may give rise to an enormous difference in the normalized RET rate. Interestingly, a very simplified model for the conductivities suggests that this disparity is predicted whenever the 2D medium near the emitters is anisotropic, as we will see in Sec. \ref{SecToyModel}. 



\subsection{RET on strained phosphorene \label{subsecB}}

\begin{figure}[b]
    \centering
    \includegraphics[width=1\linewidth]{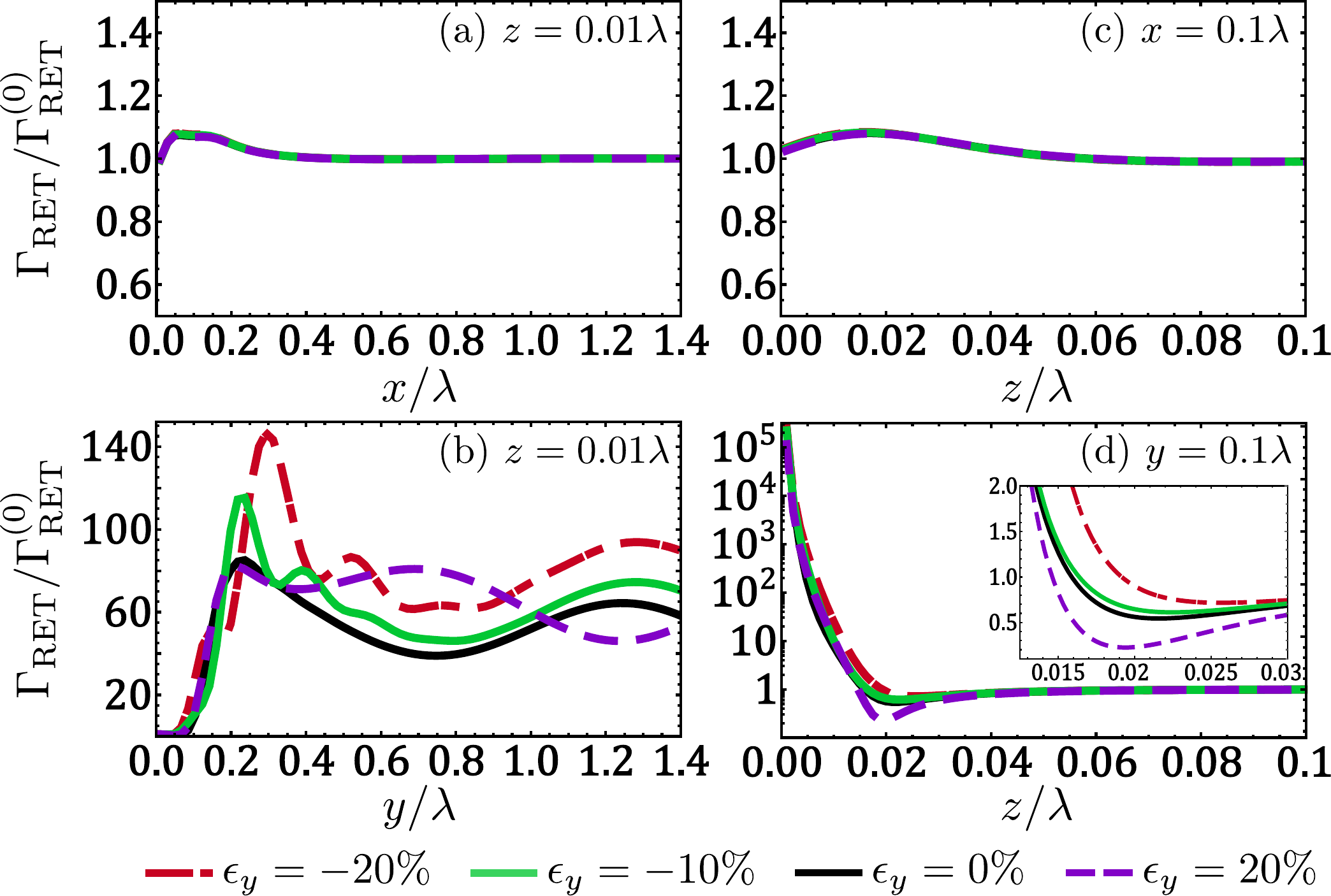}
    \caption{Normalized RET rate as a function of the distance parameters for $\lambda=10$~$\mu$m and different strain values. In the first column, we take $z=0.01\lambda$ and vary (a) $x$, with $y = 0$, and (b) $y$, with $x = 0$. For the second column, we vary the distance $z$ between the QEs and phosphorene, setting (c) $x = 0.1\lambda$, with $y = 0$, and (d) $y = 0.1\lambda$, with $x = 0$.}
    \label{Strain01}
\end{figure}

We now account for the effects of uniaxial strain in Fig.~\ref{Strain01}, evaluating the variation of the RET rate as a function of the distance parameters for different strain values. We have chosen to discuss the strain $\epsilon_y$ being applied to the zigzag direction in phosphorene, but the analysis for $\epsilon_x$ follows analogously. From Fig.~\ref{Strain01}, it is clearly seen that the RET rate is poorly influenced by strain when the QEs are separated along the $x-$axis, but it is quite sensitive when they are along the $y-$axis, according to the discussions in Sec.~\ref{subsecA}.

\begin{figure}[b!]
    \centering
    \includegraphics[width=1\linewidth]{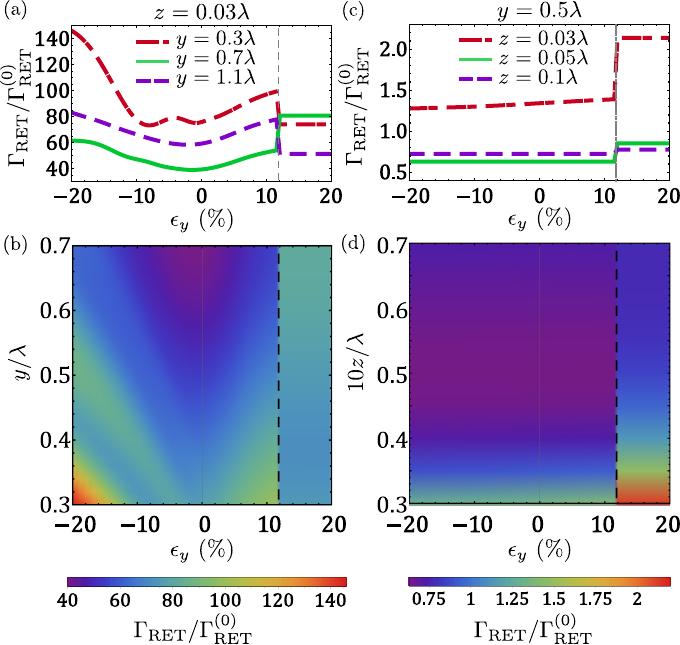}
    \caption{Normalized RET rate as a function of strain as we change the separation of the QEs. We set $z=0.03\lambda$ in panels (a) and (b), and $y = 0.5\lambda$ in panels (c) and (d). In all plots, $\lambda=10$ $\mu$m and we draw a dashed line at $\epsilon_y=11.8\%$, which separates the metallic and insulating regimes when $E_{\rm F} = 0.7$~eV.}
    \label{Fig5}
\end{figure}

Figure~\ref{Fig5} presents the modification in the RET rate as we continuously tune strain, revealing a strong non-monotonic character. In all plots, we draw dashed lines at $\epsilon_y=11.8 \%$, highlighting the strain value from which phosphorene exhibits a change from the metallic to the insulating regime (see Appendix \ref{ConductivitiesSec}). This regime, in turn, gives rise to a discontinuous change in the RET rate. From these results, we conclude that strain can be a resourceful tool for controlling RET, as it drastically alters the RET rate, especially for short distances.


\section{Anisotropic Effects on RET \label{SecToyModel}}

\begin{figure}[b]
    \centering
    \includegraphics[width=1\linewidth]{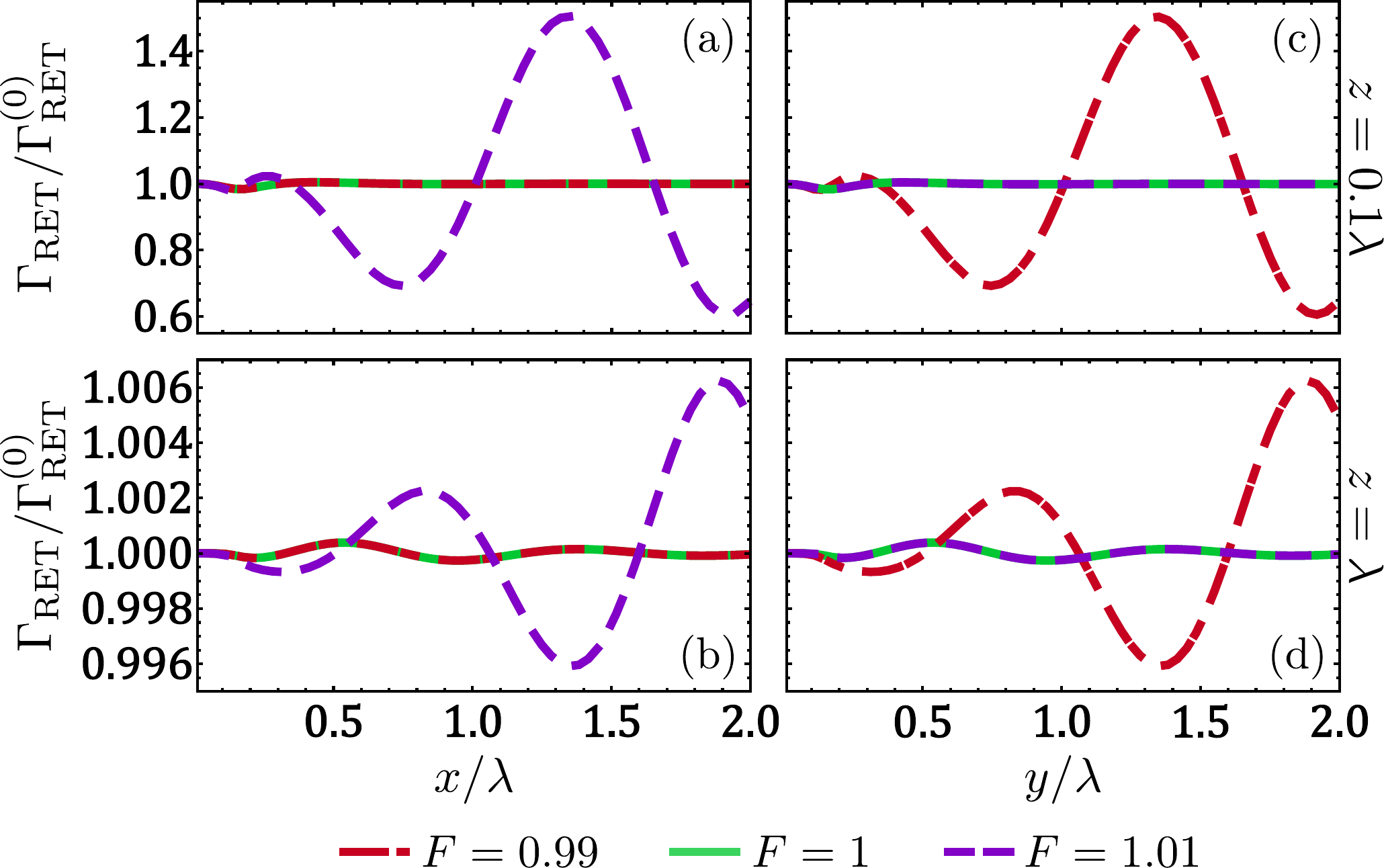}
    \caption{Normalized RET rate as a function of the distance between the QEs near anisotropic 2D material/SiC interface separated along the $x-$axis [panels (a) and (b)] and along the $y-$axis [panels (c) and (d)], for different values of $F$. We set $z = 0.1\lambda$ in the first row and $z = \lambda$ in the second one. In all cases, $\lambda=1.5$ $\mu$m.} 
    \label{Anisotropia}
\end{figure}

In Sec. \ref{Control}, we showed that the RET rate of two QEs near phosphorene remains virtually unchanged when we separate them along the armchair direction, but is drastically altered when placed along the zigzag direction. In this section, we argue that this difference is not specific to phosphorene but rather a general property of anisotropic media.

We consider a generic anisotropic 2D material grown above a SiC substrate and assume that its longitudinal conductivity along $x-$direction takes the generic value $\bar{\sigma}_{xx}=\sigma_0\left(0.33+i1.63\right)$, where $\sigma_0=e^2/\hbar$. In this toy model, we suppose that the conductivity along the $y-$direction is related to $\bar{\sigma}_{xx}$ by an anisotropy parameter $F$, i.e., $\bar{\sigma}_{yy}(F)=F\bar{\sigma}_{xx}$. The conductivity tensor reads
\begin{eqnarray}
\overleftrightarrow{\bm{\sigma}}_{\rm toy}(F)=\begin{bmatrix}
\bar{\sigma}_{xx} & 0  \\
0 & F\bar{\sigma}_{xx} 
\end{bmatrix}\,.
\label{ToyModel}
\end{eqnarray}

\noindent For $F=1$, we recover the case of an isotropic 2D material. This simple toy model allows us to isolate the impacts of the interface anisotropy on RET.

In Fig.~\ref{Anisotropia}, we portray the normalized RET rate for two emitters with $\lambda=1.5$ $\mu$m placed at a distance $z$ above the anisotropic 2D material. We explore two scenarios. In the first, the emitters are separated along the $x-$direction [Figs.~\ref{Anisotropia}(a) and \ref{Anisotropia}(b)], and, in the second case, the emitters are separated along the $y-$direction [Figs.~\ref{Anisotropia}(c) and \ref{Anisotropia}(d)]. For $F=1$, the normalized RET rate gives the same value for both scenarios, as expected for an isotropic interface. However, a slight variation in $F$ reveals that the RET rate exhibits an anisotropic character, indicating that the RET rate generally depends on the direction in which the emitters are separated. Consequently, a large variation in RET and a greater sensitivity in the parameter variation can be expected when both QEs are separated along the $x-$direction ($y-$direction) if $F>1$ ($F<1$). The conductivity model in Eq.~(\ref{ToyModel}) captures the global effects of anisotropy, which is also present in other realistic systems pertinent to nanophotonics, such as carbon nanotubes metasurfaces \cite{Rodriguez-Lopes2024} and a broad class of 2D materials \cite{Li2021}.


\section{\label{Conclusions} Final Remarks and Conclusions}

In summary, we showed that RET between QEs can be substantially enhanced or suppressed when brought close to a phosphorene sheet. We demonstrated that the anisotropy of phosphorene's optical tensors leads to a drastically different RET rate depending on whether the QEs are separated along the armchair or zigzag directions of phosphorene. In addition to its inherent anisotropic response, phosphorene's structure allows the application of uniform strain, enabling active and non-monotonical modulation of the RET rate within the scope of experimental realization. Moreover, through a simplified toy model, we verified that this anisotropic signature in the RET rate is envisioned for general anisotropic 2D media. Altogether, we believe that this anisotropy may be relevant for designing efficient energy-harvesting mechanisms in future nanophotonic devices, especially in view of the fact that other 2D materials also present an anisotropic band structure \cite{Li2021} and many of them exhibit strain-tunable optical properties. Our findings motivate further studies on strain engineering in the pool of anisotropic 2D materials and their use in nanophotonic effects.

\section*{Acknowledgments}

The authors thank the Brazilian Agencies CAPES, CNPq, and FAPERJ for financial support. J.O.-C. is grateful to CAPES (Grant No. 88887.958122/2024-00). P.P.A.
acknowledges funding from CNPq (Grant No. 152050/2024-8). C.F. acknowledges funding from CNPq and FAPERJ (Grants No. 308641/2022-1 and 204.376/2024).

\appendix

\section{\label{ConductivitiesSec} Conductivities of strained phosphorene}

In this work, we model the electronic structure of phosphorene using a two-band tight-binding model \cite{Rudenko2014, Midtvedt2017}. The effect of the uniaxial strain is included following Harrison's prescription \cite{Taghizadeh2016}. The optical conductivities along the $x-$ and $y-$directions are computed within the linear response theory and consist of two contributions: interband and intraband terms, such that $\sigma_{\mu\mu}(\omega)=\sigma_{\mu\mu}^{\mathrm{Inter}}(\omega)+\sigma_{\mu\mu}^{\mathrm{Intra}}(\omega)$, where $\mu = \{x,y\}$. A detailed review of this approach is provided in Ref.~\cite{Patricia2024}.

\begin{figure}[t]
    \centering
    \includegraphics[width=0.95\linewidth]{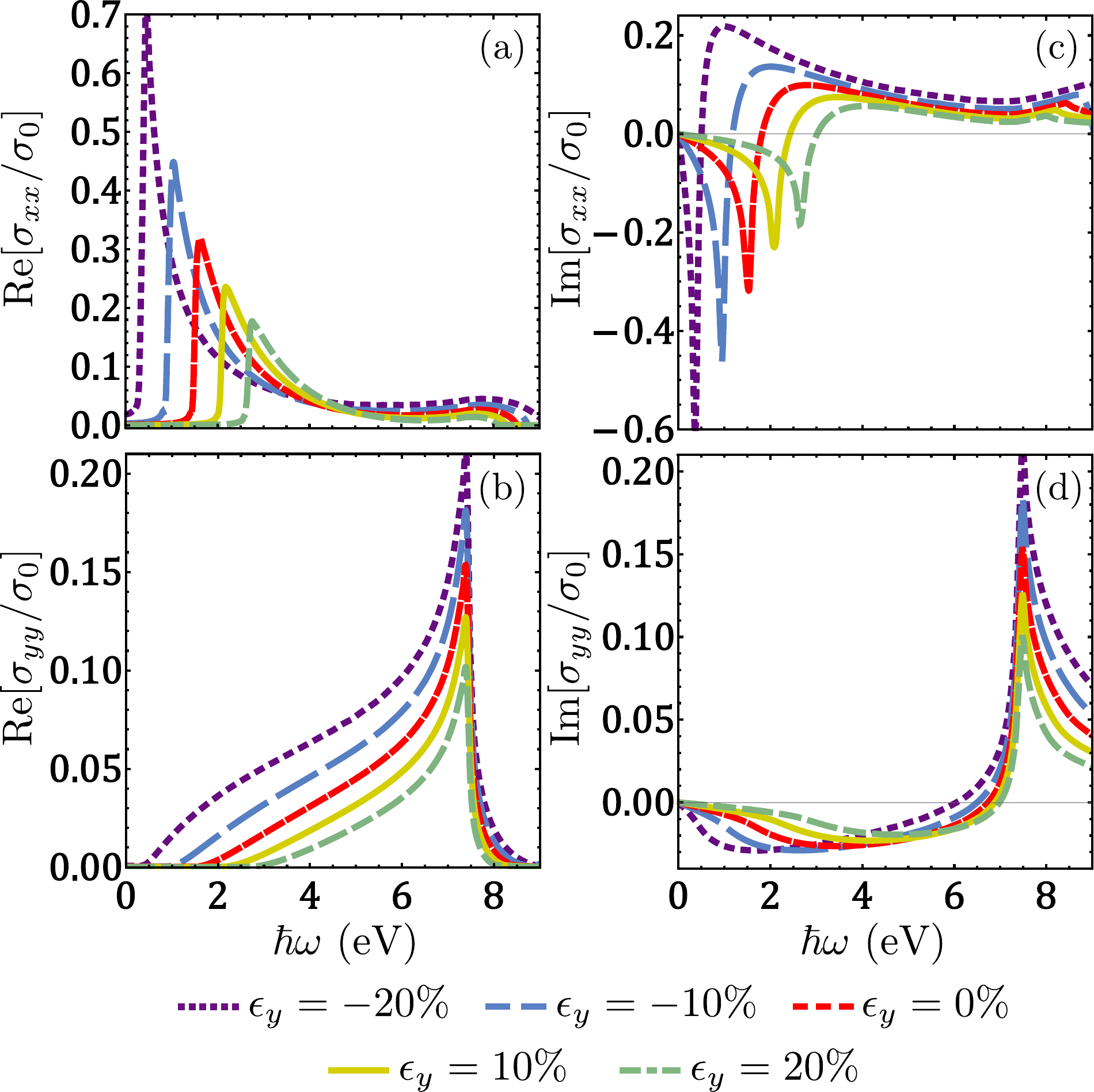}
    \caption{Real and imaginary parts of phosphorene’s optical conductivity in the insulator regime ($E_{\rm F}$ always inside the energy gap) as functions of frequency under the effect of different strains.}
    \label{Conductivities}
\end{figure}

\begin{figure}[t!]
    \centering
    \includegraphics[width=0.95\linewidth]{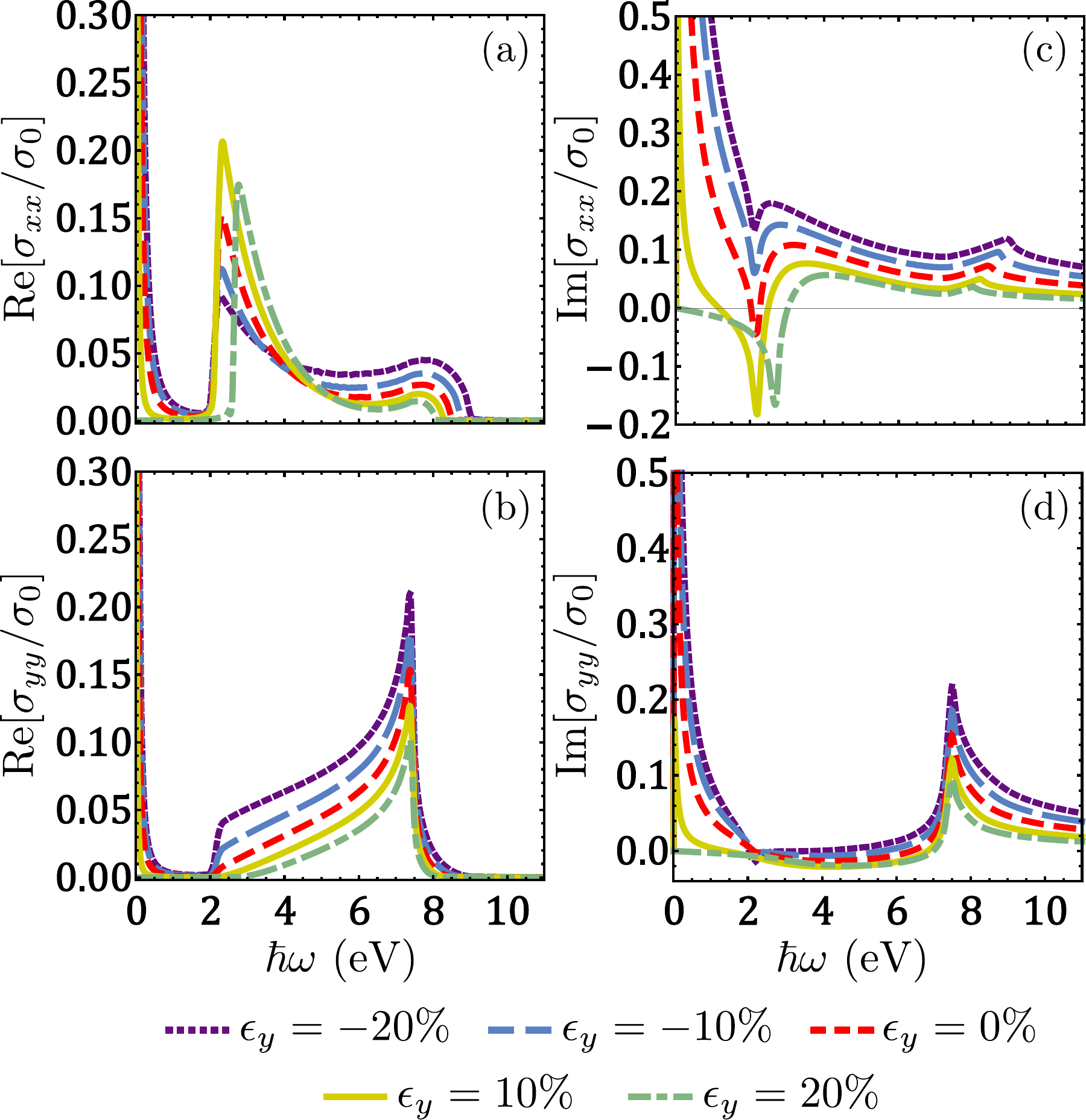}
    \caption{Real and imaginary parts of phosphorene’s optical conductivity for $E_{\rm F} = 0.7$~eV as functions of frequency under the effect of different strains.}
    \label{Conductivities2}
\end{figure}

In Figs.~\ref{Conductivities} and \ref{Conductivities2}, we show the real and imaginary parts of the optical conductivity of phosphorene under different values of uniaxial strain $\epsilon_y$ applied along the $y-$direction. In Fig.~\ref{Conductivities}, we display the results where the Fermi energy $E_{\rm F}$ is forced to be in the middle of the energy bandgap, and only the interband term contributes to the optical conductivity. In Fig.~\ref{Conductivities2}, we display the results for $E_{\rm F}=0.7$~eV. In this situation, a Drude peak at low frequencies, associated with the intraband contribution, appears when $E_{\rm F}$ crosses an electronic energy band of phosphorene (for $E_{\rm F}=0.7$~eV, the Drude peak occurs when $\epsilon_y<11.8\%$) \cite{Patricia2024}. In the main text, we use the conductivities reported in Fig.~\ref{Conductivities2}.

\section{\label{Fresnel} Strained phosphorene reflection coefficients}

To enable the application of uniaxial strain on phosphorene, we assume that it is grown on a slab of SiC, whose electrical permittivity can be described by a simple Drude-Lorentz model \cite{Wald}, such that
\begin{equation}
    \varepsilon_{\mathrm{SiC}}(\omega)=\varepsilon_\infty\left(1+\dfrac{\omega_L^2-\omega_T^2}{\omega_T^2-\omega^2-i\omega\gamma_{\mathrm{SiC}}}\right)\,,
\end{equation}

\noindent where $\varepsilon_\infty=6.7\varepsilon_0$, $\omega_L=182.7\times10^{12}$~rad/s, $\omega_T=149.5\times10^{12}$~rad/s and $\gamma_{\mathrm{SiC}}=0.9\times10^{12}$~rad/s \cite{palik1998}.

The Fresnel reflection coefficients of the phosphorene/SiC interface
\begin{align}
    r^{p,p}({\bm k}_{\parallel},\omega,\epsilon_{\mu})&=\dfrac{\Delta^{T}_+\Delta^L_-+\Lambda^2}{\Delta^{T}_+\Delta^L_++\Lambda^2}\, ,\\
    r^{s,s}({\bm k}_{\parallel},\omega,\epsilon_{\mu})&=-\dfrac{\Delta^{T}_-\Delta^L_++\Lambda^2}{\Delta^{T}_+\Delta^L_++\Lambda^2}
    \label{fresnel}
\end{align}

\noindent can be easily obtained by solving Maxwell's equations with appropriate boundary conditions \cite{KortKamp2015}, where
\begin{align}
\Delta_{{\pm}}^T({\bm k}_{\parallel},\omega,\epsilon_{\mu})&=\dfrac{1}{\mu_0}\bigg[k_{z2}({\bm k}_{\parallel},\omega)\mu_1\pm k_{z1}({\bm k}_{\parallel},\omega)\mu_1 \nonumber \\
&+ \,\omega \mu_1\mu_2\sigma_T({\bm k}_{\parallel},\omega,\epsilon_{\mu})\bigg], \\
    \Delta_{{\pm}}^L({\bm k}_{\parallel},\omega,\epsilon_{\mu})&=\dfrac{1}{\varepsilon_0}\bigg[k_{z1}({\bm k}_{\parallel}, \omega) \varepsilon_2 \pm k_{z2}({\bm k}_{\parallel},\omega)\varepsilon_1 \nonumber \\
    &+\,k_{z1}({\bm k}_{\parallel},\omega)k_{z2}({\bm k}_{\parallel},\omega)\dfrac{\sigma_L({\bm k}_{\parallel},\omega,\epsilon_{\mu})}{\omega}\bigg],\\
    \Lambda^2({\bm k}_{\parallel},\omega,\epsilon_{\mu})&=-\dfrac{\mu_1\mu_2}{\varepsilon_0^2}k_{z1}({\bm k}_{\parallel},\omega)k_{z2}({\bm k}_{\parallel},\omega)\sigma_{LT}^2({\bm k}_{\parallel},\omega,\epsilon_{\mu}) \nonumber\,. \\
\end{align}

\noindent In the previous equations, $\epsilon_\mu$ is the uniaxial strain applied in phosphorene along the $\mu-$direction, $\sigma_L=(k_x^2\sigma_{xx}+k_y^2\sigma_{yy})/k_{\parallel}^2$, $\sigma_T=(k_y^2\sigma_{xx}+k_x^2\sigma_{yy})/k_{\parallel}^2$, and $\sigma_{LT}=k_xk_y(\sigma_{yy}-\sigma_{xx})/k_{\parallel}^2$. Additionally, $k_{zj}=\sqrt{\omega^2\epsilon_j\mu_j-k_{\parallel}^2}$. For our system, the first medium is vacuum ($\varepsilon_1=\varepsilon_0$, $\mu_1=\mu_0$) and the second is SiC ($\varepsilon_2=\varepsilon_\mathrm{SiC}(\omega)$, $\mu_2=\mu_0$).

\providecommand{\noopsort}[1]{}\providecommand{\singleletter}[1]{#1}%
%


\begin{thebibliography}{87}%
\makeatletter
\providecommand \@ifxundefined [1]{%
 \@ifx{#1\undefined}
}%
\providecommand \@ifnum [1]{%
 \ifnum #1\expandafter \@firstoftwo
 \else \expandafter \@secondoftwo
 \fi
}%
\providecommand \@ifx [1]{%
 \ifx #1\expandafter \@firstoftwo
 \else \expandafter \@secondoftwo
 \fi
}%
\providecommand \natexlab [1]{#1}%
\providecommand \enquote  [1]{``#1''}%
\providecommand \bibnamefont  [1]{#1}%
\providecommand \bibfnamefont [1]{#1}%
\providecommand \citenamefont [1]{#1}%
\providecommand \href@noop [0]{\@secondoftwo}%
\providecommand \href [0]{\begingroup \@sanitize@url \@href}%
\providecommand \@href[1]{\@@startlink{#1}\@@href}%
\providecommand \@@href[1]{\endgroup#1\@@endlink}%
\providecommand \@sanitize@url [0]{\catcode `\\12\catcode `\$12\catcode `\&12\catcode `\#12\catcode `\^12\catcode `\_12\catcode `\%12\relax}%
\providecommand \@@startlink[1]{}%
\providecommand \@@endlink[0]{}%
\providecommand \url  [0]{\begingroup\@sanitize@url \@url }%
\providecommand \@url [1]{\endgroup\@href {#1}{\urlprefix }}%
\providecommand \urlprefix  [0]{URL }%
\providecommand \Eprint [0]{\href }%
\providecommand \doibase [0]{https://doi.org/}%
\providecommand \selectlanguage [0]{\@gobble}%
\providecommand \bibinfo  [0]{\@secondoftwo}%
\providecommand \bibfield  [0]{\@secondoftwo}%
\providecommand \translation [1]{[#1]}%
\providecommand \BibitemOpen [0]{}%
\providecommand \bibitemStop [0]{}%
\providecommand \bibitemNoStop [0]{.\EOS\space}%
\providecommand \EOS [0]{\spacefactor3000\relax}%
\providecommand \BibitemShut  [1]{\csname bibitem#1\endcsname}%
\let\auto@bib@innerbib\@empty
\bibitem [{\citenamefont {Morgner}\ \emph {et~al.}(2023)\citenamefont {Morgner}, \citenamefont {Tu}, \citenamefont {K{\"o}nig}, \citenamefont {Sailer}, \citenamefont {Hei{\ss}e}, \citenamefont {Bekker}, \citenamefont {Sikora}, \citenamefont {Lyu}, \citenamefont {Yerokhin}, \citenamefont {Harman} \emph {et~al.}}]{Morgner2023}%
  \BibitemOpen
  \bibfield  {author} {\bibinfo {author} {\bibfnamefont {J.}~\bibnamefont {Morgner}}, \bibinfo {author} {\bibfnamefont {B.}~\bibnamefont {Tu}}, \bibinfo {author} {\bibfnamefont {C.}~\bibnamefont {K{\"o}nig}}, \bibinfo {author} {\bibfnamefont {T.}~\bibnamefont {Sailer}}, \bibinfo {author} {\bibfnamefont {F.}~\bibnamefont {Hei{\ss}e}}, \bibinfo {author} {\bibfnamefont {H.}~\bibnamefont {Bekker}}, \bibinfo {author} {\bibfnamefont {B.}~\bibnamefont {Sikora}}, \bibinfo {author} {\bibfnamefont {C.}~\bibnamefont {Lyu}}, \bibinfo {author} {\bibfnamefont {V.}~\bibnamefont {Yerokhin}}, \bibinfo {author} {\bibfnamefont {Z.}~\bibnamefont {Harman}}, \emph {et~al.},\ }\bibfield  {title} {\bibinfo {title} {Stringent test of {QED} with hydrogen-like tin},\ }\href {https://doi.org/10.1038/s41586-023-06453-2} {\bibfield  {journal} {\bibinfo  {journal} {Nature}\ }\textbf {\bibinfo {volume} {622}},\ \bibinfo {pages} {53} (\bibinfo {year} {2023})}\BibitemShut {NoStop}%
\bibitem [{\citenamefont {Aguillard}\ \emph {et~al.}(2024)\citenamefont {Aguillard}, \citenamefont {Albahri}, \citenamefont {Allspach}, \citenamefont {Anisenkov}, \citenamefont {Badgley}, \citenamefont {Bae{\ss}ler}, \citenamefont {Bailey}, \citenamefont {Bailey}, \citenamefont {Baranov}, \citenamefont {Barlas-Yucel} \emph {et~al.}}]{Aguillard2024}%
  \BibitemOpen
  \bibfield  {author} {\bibinfo {author} {\bibfnamefont {D.}~\bibnamefont {Aguillard}}, \bibinfo {author} {\bibfnamefont {T.}~\bibnamefont {Albahri}}, \bibinfo {author} {\bibfnamefont {D.}~\bibnamefont {Allspach}}, \bibinfo {author} {\bibfnamefont {A.}~\bibnamefont {Anisenkov}}, \bibinfo {author} {\bibfnamefont {K.}~\bibnamefont {Badgley}}, \bibinfo {author} {\bibfnamefont {S.}~\bibnamefont {Bae{\ss}ler}}, \bibinfo {author} {\bibfnamefont {I.}~\bibnamefont {Bailey}}, \bibinfo {author} {\bibfnamefont {L.}~\bibnamefont {Bailey}}, \bibinfo {author} {\bibfnamefont {V.}~\bibnamefont {Baranov}}, \bibinfo {author} {\bibfnamefont {E.}~\bibnamefont {Barlas-Yucel}}, \emph {et~al.},\ }\bibfield  {title} {\bibinfo {title} {Detailed report on the measurement of the positive muon anomalous magnetic moment to 0.20 ppm},\ }\href {http://dx.doi.org/10.1103/PhysRevD.110.032009} {\bibfield  {journal} {\bibinfo  {journal} {Physical Review D}\ }\textbf {\bibinfo {volume} {110}},\ \bibinfo {pages} {032009} (\bibinfo {year}
  {2024})}\BibitemShut {NoStop}%
\bibitem [{\citenamefont {Loetzsch}\ \emph {et~al.}(2024)\citenamefont {Loetzsch}, \citenamefont {Beyer}, \citenamefont {Duval}, \citenamefont {Spillmann}, \citenamefont {Bana{\'s}}, \citenamefont {Dergham}, \citenamefont {Kr{\"o}ger}, \citenamefont {Glorius}, \citenamefont {Grisenti}, \citenamefont {Guerra} \emph {et~al.}}]{Loetzsch2024}%
  \BibitemOpen
  \bibfield  {author} {\bibinfo {author} {\bibfnamefont {R.}~\bibnamefont {Loetzsch}}, \bibinfo {author} {\bibfnamefont {H.}~\bibnamefont {Beyer}}, \bibinfo {author} {\bibfnamefont {L.}~\bibnamefont {Duval}}, \bibinfo {author} {\bibfnamefont {U.}~\bibnamefont {Spillmann}}, \bibinfo {author} {\bibfnamefont {D.}~\bibnamefont {Bana{\'s}}}, \bibinfo {author} {\bibfnamefont {P.}~\bibnamefont {Dergham}}, \bibinfo {author} {\bibfnamefont {F.}~\bibnamefont {Kr{\"o}ger}}, \bibinfo {author} {\bibfnamefont {J.}~\bibnamefont {Glorius}}, \bibinfo {author} {\bibfnamefont {R.}~\bibnamefont {Grisenti}}, \bibinfo {author} {\bibfnamefont {M.}~\bibnamefont {Guerra}}, \emph {et~al.},\ }\bibfield  {title} {\bibinfo {title} {Testing quantum electrodynamics in extreme fields using helium-like uranium},\ }\href {http://dx.doi.org/10.1038/s41586-023-06910-y} {\bibfield  {journal} {\bibinfo  {journal} {Nature}\ }\textbf {\bibinfo {volume} {625}},\ \bibinfo {pages} {673} (\bibinfo {year} {2024})}\BibitemShut {NoStop}%
\bibitem [{\citenamefont {Milonni}(2013)}]{milonni2013}%
  \BibitemOpen
  \bibfield  {author} {\bibinfo {author} {\bibfnamefont {P.}~\bibnamefont {Milonni}},\ }\href {https://books.google.com.br/books?id=uPHJCgAAQBAJ} {\emph {\bibinfo {title} {The Quantum Vacuum}}}\ (\bibinfo  {publisher} {Academic Press},\ \bibinfo {year} {2013})\BibitemShut {NoStop}%
\bibitem [{\citenamefont {Milton}(2001)}]{milton2001casimir}%
  \BibitemOpen
  \bibfield  {author} {\bibinfo {author} {\bibfnamefont {K.}~\bibnamefont {Milton}},\ }\href {https://books.google.com.br/books?id=s07VCgAAQBAJ} {\emph {\bibinfo {title} {Casimir Effect}}}\ (\bibinfo  {publisher} {World Scientific Publishing Company},\ \bibinfo {year} {2001})\BibitemShut {NoStop}%
\bibitem [{\citenamefont {Dalvit}\ \emph {et~al.}(2011)\citenamefont {Dalvit}, \citenamefont {Milonni}, \citenamefont {Roberts},\ and\ \citenamefont {da~Rosa}}]{dalvit2011}%
  \BibitemOpen
  \bibfield  {author} {\bibinfo {author} {\bibfnamefont {D.}~\bibnamefont {Dalvit}}, \bibinfo {author} {\bibfnamefont {P.}~\bibnamefont {Milonni}}, \bibinfo {author} {\bibfnamefont {D.}~\bibnamefont {Roberts}},\ and\ \bibinfo {author} {\bibfnamefont {F.}~\bibnamefont {da~Rosa}},\ }\href {https://books.google.com.br/books?id=eqQRBwAAQBAJ} {\emph {\bibinfo {title} {Casimir Physics}}},\ Lecture Notes in Physics\ (\bibinfo  {publisher} {Springer Berlin Heidelberg},\ \bibinfo {year} {2011})\BibitemShut {NoStop}%
\bibitem [{\citenamefont {Novotny}\ and\ \citenamefont {Hecht}(2012)}]{Novotny_Hecht_2012}%
  \BibitemOpen
  \bibfield  {author} {\bibinfo {author} {\bibfnamefont {L.}~\bibnamefont {Novotny}}\ and\ \bibinfo {author} {\bibfnamefont {B.}~\bibnamefont {Hecht}},\ }\href@noop {} {\emph {\bibinfo {title} {Principles of nano-optics}}}\ (\bibinfo  {publisher} {Cambridge university press},\ \bibinfo {year} {2012})\BibitemShut {NoStop}%
\bibitem [{\citenamefont {Dodonov}(2020)}]{dodonov2020}%
  \BibitemOpen
  \bibfield  {author} {\bibinfo {author} {\bibfnamefont {V.}~\bibnamefont {Dodonov}},\ }\bibfield  {title} {\bibinfo {title} {Fifty years of the dynamical {C}asimir effect},\ }\href {https://www.mdpi.com/2624-8174/2/1/7} {\bibfield  {journal} {\bibinfo  {journal} {Physics}\ }\textbf {\bibinfo {volume} {2}},\ \bibinfo {pages} {67} (\bibinfo {year} {2020})}\BibitemShut {NoStop}%
\bibitem [{\citenamefont {Andrews}\ and\ \citenamefont {Sherborne}(1987)}]{Andres1987}%
  \BibitemOpen
  \bibfield  {author} {\bibinfo {author} {\bibfnamefont {D.~L.}\ \bibnamefont {Andrews}}\ and\ \bibinfo {author} {\bibfnamefont {B.~S.}\ \bibnamefont {Sherborne}},\ }\bibfield  {title} {\bibinfo {title} {Resonant excitation transfer: A quantum electrodynamical study},\ }\href {https://doi.org/10.1063/1.451910} {\bibfield  {journal} {\bibinfo  {journal} {The Journal of Chemical Physics}\ }\textbf {\bibinfo {volume} {86}},\ \bibinfo {pages} {4011} (\bibinfo {year} {1987})}\BibitemShut {NoStop}%
\bibitem [{\citenamefont {Jones}\ and\ \citenamefont {Bradshaw}(2019)}]{Jones2019}%
  \BibitemOpen
  \bibfield  {author} {\bibinfo {author} {\bibfnamefont {G.}~\bibnamefont {Jones}}\ and\ \bibinfo {author} {\bibfnamefont {D.}~\bibnamefont {Bradshaw}},\ }\bibfield  {title} {\bibinfo {title} {Resonance energy transfer: From fundamental theory to recent applications},\ }\href {https://doi.org/10.3389/fphy.2019.00100} {\bibfield  {journal} {\bibinfo  {journal} {Frontiers in Physics}\ }\textbf {\bibinfo {volume} {7}},\ \bibinfo {pages} {100} (\bibinfo {year} {2019})}\BibitemShut {NoStop}%
\bibitem [{\citenamefont {Andrews}\ and\ \citenamefont {Lipson}(2021)}]{Andrews2021}%
  \BibitemOpen
  \bibfield  {author} {\bibinfo {author} {\bibfnamefont {D.~L.}\ \bibnamefont {Andrews}}\ and\ \bibinfo {author} {\bibfnamefont {R.~H.}\ \bibnamefont {Lipson}},\ }\href {https://doi.org/10.1088/978-0-7503-3683-3} {\emph {\bibinfo {title} {Molecular Photophysics and Spectroscopy}}}\ (\bibinfo  {publisher} {IOP Publishing},\ \bibinfo {year} {2021})\BibitemShut {NoStop}%
\bibitem [{\citenamefont {Förster}(2012)}]{Forster1946}%
  \BibitemOpen
  \bibfield  {author} {\bibinfo {author} {\bibfnamefont {T.}~\bibnamefont {Förster}},\ }\bibfield  {title} {\bibinfo {title} {Energy migration and fluorescence},\ }\href {https://doi.org/10.1117/1.JBO.17.1.011002} {\bibfield  {journal} {\bibinfo  {journal} {Journal of Biomedical Optics}\ }\textbf {\bibinfo {volume} {17}},\ \bibinfo {pages} {011002} (\bibinfo {year} {2012})},\ \bibinfo {note} {originally published in 1946}\BibitemShut {NoStop}%
\bibitem [{\citenamefont {Chen}\ \emph {et~al.}(2012)\citenamefont {Chen}, \citenamefont {Cheng}, \citenamefont {Liu}, \citenamefont {Souris}, \citenamefont {Chen}, \citenamefont {Mou},\ and\ \citenamefont {Lo}}]{Chen2012}%
  \BibitemOpen
  \bibfield  {author} {\bibinfo {author} {\bibfnamefont {N.-T.}\ \bibnamefont {Chen}}, \bibinfo {author} {\bibfnamefont {S.-H.}\ \bibnamefont {Cheng}}, \bibinfo {author} {\bibfnamefont {C.-P.}\ \bibnamefont {Liu}}, \bibinfo {author} {\bibfnamefont {J.~S.}\ \bibnamefont {Souris}}, \bibinfo {author} {\bibfnamefont {C.-T.}\ \bibnamefont {Chen}}, \bibinfo {author} {\bibfnamefont {C.-Y.}\ \bibnamefont {Mou}},\ and\ \bibinfo {author} {\bibfnamefont {L.-W.}\ \bibnamefont {Lo}},\ }\bibfield  {title} {\bibinfo {title} {Recent advances in nanoparticle-based {F}örster resonance energy transfer for biosensing, molecular imaging and drug release profiling},\ }\href {https://www.mdpi.com/1422-0067/13/12/16598} {\bibfield  {journal} {\bibinfo  {journal} {International Journal of Molecular Sciences}\ }\textbf {\bibinfo {volume} {13}},\ \bibinfo {pages} {16598} (\bibinfo {year} {2012})}\BibitemShut {NoStop}%
\bibitem [{\citenamefont {Geißler}\ \emph {et~al.}(2013)\citenamefont {Geißler}, \citenamefont {Stufler}, \citenamefont {L{\"o}hmannsr{\"o}ben},\ and\ \citenamefont {Hildebrandt}}]{Geißler2013}%
  \BibitemOpen
  \bibfield  {author} {\bibinfo {author} {\bibfnamefont {D.}~\bibnamefont {Geißler}}, \bibinfo {author} {\bibfnamefont {S.}~\bibnamefont {Stufler}}, \bibinfo {author} {\bibfnamefont {H.-G.}\ \bibnamefont {L{\"o}hmannsr{\"o}ben}},\ and\ \bibinfo {author} {\bibfnamefont {N.}~\bibnamefont {Hildebrandt}},\ }\bibfield  {title} {\bibinfo {title} {Six-color time-resolved {F}örster resonance energy transfer for ultrasensitive multiplexed biosensing},\ }\href {https://pubs.acs.org/doi/abs/10.1021/ja310317n} {\bibfield  {journal} {\bibinfo  {journal} {Journal of the American Chemical Society}\ }\textbf {\bibinfo {volume} {135}},\ \bibinfo {pages} {1102} (\bibinfo {year} {2013})}\BibitemShut {NoStop}%
\bibitem [{\citenamefont {Hussain}\ \emph {et~al.}(2014)\citenamefont {Hussain}, \citenamefont {Dey}, \citenamefont {Chakraborty}, \citenamefont {Saha}, \citenamefont {Datta~Roy}, \citenamefont {Chakraborty}, \citenamefont {Debnath},\ and\ \citenamefont {Bhattacharjee}}]{Hussain2014}%
  \BibitemOpen
  \bibfield  {author} {\bibinfo {author} {\bibfnamefont {S.~A.}\ \bibnamefont {Hussain}}, \bibinfo {author} {\bibfnamefont {D.}~\bibnamefont {Dey}}, \bibinfo {author} {\bibfnamefont {S.}~\bibnamefont {Chakraborty}}, \bibinfo {author} {\bibfnamefont {J.}~\bibnamefont {Saha}}, \bibinfo {author} {\bibfnamefont {A.}~\bibnamefont {Datta~Roy}}, \bibinfo {author} {\bibfnamefont {S.}~\bibnamefont {Chakraborty}}, \bibinfo {author} {\bibfnamefont {P.}~\bibnamefont {Debnath}},\ and\ \bibinfo {author} {\bibfnamefont {D.}~\bibnamefont {Bhattacharjee}},\ }\bibfield  {title} {\bibinfo {title} {Fluorescence resonance energy transfer ({FRET}) sensor},\ }\href@noop {} {\bibfield  {journal} {\bibinfo  {journal} {Journal of Spectroscopy and Dynamics}\ } (\bibinfo {year} {2014})}\BibitemShut {NoStop}%
\bibitem [{\citenamefont {Yang}\ \emph {et~al.}(2016)\citenamefont {Yang}, \citenamefont {Cui}, \citenamefont {Wang}, \citenamefont {Lei},\ and\ \citenamefont {Zhang}}]{Yang2016}%
  \BibitemOpen
  \bibfield  {author} {\bibinfo {author} {\bibfnamefont {L.}~\bibnamefont {Yang}}, \bibinfo {author} {\bibfnamefont {C.}~\bibnamefont {Cui}}, \bibinfo {author} {\bibfnamefont {L.}~\bibnamefont {Wang}}, \bibinfo {author} {\bibfnamefont {J.}~\bibnamefont {Lei}},\ and\ \bibinfo {author} {\bibfnamefont {J.}~\bibnamefont {Zhang}},\ }\bibfield  {title} {\bibinfo {title} {Dual-shell fluorescent nanoparticles for self-monitoring of p{H}-responsive molecule-releasing in a visualized way},\ }\href {https://pubs.acs.org/doi/abs/10.1021/acsami.6b05872} {\bibfield  {journal} {\bibinfo  {journal} {ACS Applied Materials \& Interfaces}\ }\textbf {\bibinfo {volume} {8}},\ \bibinfo {pages} {19084} (\bibinfo {year} {2016})}\BibitemShut {NoStop}%
\bibitem [{\citenamefont {Verma}\ \emph {et~al.}(2023)\citenamefont {Verma}, \citenamefont {Noumani}, \citenamefont {Yadav},\ and\ \citenamefont {Solanki}}]{Verma2023}%
  \BibitemOpen
  \bibfield  {author} {\bibinfo {author} {\bibfnamefont {A.~K.}\ \bibnamefont {Verma}}, \bibinfo {author} {\bibfnamefont {A.}~\bibnamefont {Noumani}}, \bibinfo {author} {\bibfnamefont {A.~K.}\ \bibnamefont {Yadav}},\ and\ \bibinfo {author} {\bibfnamefont {P.~R.}\ \bibnamefont {Solanki}},\ }\bibfield  {title} {\bibinfo {title} {{FRET} based biosensor: Principle applications recent advances and challenges},\ }\href {https://doi.org/10.3390/diagnostics13081375} {\bibfield  {journal} {\bibinfo  {journal} {Diagnostics (Basel)}\ }\textbf {\bibinfo {volume} {13}},\ \bibinfo {pages} {1375} (\bibinfo {year} {2023})}\BibitemShut {NoStop}%
\bibitem [{\citenamefont {Ha}\ \emph {et~al.}(2024)\citenamefont {Ha}, \citenamefont {Fei}, \citenamefont {Schmid}, \citenamefont {Lee}, \citenamefont {Gonzalez~Jr}, \citenamefont {Paul},\ and\ \citenamefont {Yeou}}]{Ha2024}%
  \BibitemOpen
  \bibfield  {author} {\bibinfo {author} {\bibfnamefont {T.}~\bibnamefont {Ha}}, \bibinfo {author} {\bibfnamefont {J.}~\bibnamefont {Fei}}, \bibinfo {author} {\bibfnamefont {S.}~\bibnamefont {Schmid}}, \bibinfo {author} {\bibfnamefont {N.~K.}\ \bibnamefont {Lee}}, \bibinfo {author} {\bibfnamefont {R.~L.}\ \bibnamefont {Gonzalez~Jr}}, \bibinfo {author} {\bibfnamefont {S.}~\bibnamefont {Paul}},\ and\ \bibinfo {author} {\bibfnamefont {S.}~\bibnamefont {Yeou}},\ }\bibfield  {title} {\bibinfo {title} {Fluorescence resonance energy transfer at the single-molecule level},\ }\href {https://doi.org/10.1038/s43586-024-00298-3} {\bibfield  {journal} {\bibinfo  {journal} {Nature Reviews Methods Primers}\ }\textbf {\bibinfo {volume} {4}},\ \bibinfo {pages} {21} (\bibinfo {year} {2024})}\BibitemShut {NoStop}%
\bibitem [{\citenamefont {Sahoo}(2011)}]{Sahoo2011}%
  \BibitemOpen
  \bibfield  {author} {\bibinfo {author} {\bibfnamefont {H.}~\bibnamefont {Sahoo}},\ }\bibfield  {title} {\bibinfo {title} {Förster resonance energy transfer -- a spectroscopic nanoruler: Principle and applications},\ }\href {https://doi.org/https://doi.org/10.1016/j.jphotochemrev.2011.05.001} {\bibfield  {journal} {\bibinfo  {journal} {Journal of Photochemistry and Photobiology C: Photochemistry Reviews}\ }\textbf {\bibinfo {volume} {12}},\ \bibinfo {pages} {20} (\bibinfo {year} {2011})}\BibitemShut {NoStop}%
\bibitem [{\citenamefont {Baryshnikova}\ \emph {et~al.}(2015)\citenamefont {Baryshnikova}, \citenamefont {Petrov},\ and\ \citenamefont {Vartanyan}}]{Baryshnikova2015}%
  \BibitemOpen
  \bibfield  {author} {\bibinfo {author} {\bibfnamefont {K.~V.}\ \bibnamefont {Baryshnikova}}, \bibinfo {author} {\bibfnamefont {M.~I.}\ \bibnamefont {Petrov}},\ and\ \bibinfo {author} {\bibfnamefont {T.~A.}\ \bibnamefont {Vartanyan}},\ }\bibfield  {title} {\bibinfo {title} {Plasmon nanoruler for monitoring of transient interactions},\ }\href {https://doi.org/https://doi.org/10.1002/pssr.201510330} {\bibfield  {journal} {\bibinfo  {journal} {physica status solidi (RRL) – Rapid Research Letters}\ }\textbf {\bibinfo {volume} {9}},\ \bibinfo {pages} {711} (\bibinfo {year} {2015})}\BibitemShut {NoStop}%
\bibitem [{\citenamefont {Zhang}\ \emph {et~al.}(2023)\citenamefont {Zhang}, \citenamefont {Fang}, \citenamefont {Huang}, \citenamefont {Li}, \citenamefont {Jiang}, \citenamefont {Wang},\ and\ \citenamefont {Liu}}]{Zhang2023}%
  \BibitemOpen
  \bibfield  {author} {\bibinfo {author} {\bibfnamefont {Y.}~\bibnamefont {Zhang}}, \bibinfo {author} {\bibfnamefont {X.}~\bibnamefont {Fang}}, \bibinfo {author} {\bibfnamefont {W.}~\bibnamefont {Huang}}, \bibinfo {author} {\bibfnamefont {Q.}~\bibnamefont {Li}}, \bibinfo {author} {\bibfnamefont {H.}~\bibnamefont {Jiang}}, \bibinfo {author} {\bibfnamefont {C.}~\bibnamefont {Wang}},\ and\ \bibinfo {author} {\bibfnamefont {H.}~\bibnamefont {Liu}},\ }\bibfield  {title} {\bibinfo {title} {Plasmon resonance energy transfer nanoruler for pinpointing molecular distance and interaction on the living cell membrane},\ }\href {https://pubs.acs.org/doi/abs/10.1021/acs.nanolett.3c01629} {\bibfield  {journal} {\bibinfo  {journal} {Nano Letters}\ }\textbf {\bibinfo {volume} {23}},\ \bibinfo {pages} {7750} (\bibinfo {year} {2023})}\BibitemShut {NoStop}%
\bibitem [{\citenamefont {Teunissen}\ \emph {et~al.}(2018)\citenamefont {Teunissen}, \citenamefont {Pérez-Medina}, \citenamefont {Meijerink},\ and\ \citenamefont {Mulder}}]{Teunissen2018}%
  \BibitemOpen
  \bibfield  {author} {\bibinfo {author} {\bibfnamefont {A.~J.~P.}\ \bibnamefont {Teunissen}}, \bibinfo {author} {\bibfnamefont {C.}~\bibnamefont {Pérez-Medina}}, \bibinfo {author} {\bibfnamefont {A.}~\bibnamefont {Meijerink}},\ and\ \bibinfo {author} {\bibfnamefont {W.~J.~M.}\ \bibnamefont {Mulder}},\ }\bibfield  {title} {\bibinfo {title} {Investigating supramolecular systems using {F}örster resonance energy transfer},\ }\href {https://doi.org/10.1039/C8CS00278A} {\bibfield  {journal} {\bibinfo  {journal} {Chemical Society Reviews}\ }\textbf {\bibinfo {volume} {47}},\ \bibinfo {pages} {7027} (\bibinfo {year} {2018})}\BibitemShut {NoStop}%
\bibitem [{\citenamefont {Mayoral}\ \emph {et~al.}(2018)\citenamefont {Mayoral}, \citenamefont {Serrano-Molina}, \citenamefont {Camacho-García}, \citenamefont {Magdalena-Estirado}, \citenamefont {Blanco-Lomas}, \citenamefont {Fadaei},\ and\ \citenamefont {González-Rodríguez}}]{Mayoral2018}%
  \BibitemOpen
  \bibfield  {author} {\bibinfo {author} {\bibfnamefont {M.~J.}\ \bibnamefont {Mayoral}}, \bibinfo {author} {\bibfnamefont {D.}~\bibnamefont {Serrano-Molina}}, \bibinfo {author} {\bibfnamefont {J.}~\bibnamefont {Camacho-García}}, \bibinfo {author} {\bibfnamefont {E.}~\bibnamefont {Magdalena-Estirado}}, \bibinfo {author} {\bibfnamefont {M.}~\bibnamefont {Blanco-Lomas}}, \bibinfo {author} {\bibfnamefont {E.}~\bibnamefont {Fadaei}},\ and\ \bibinfo {author} {\bibfnamefont {D.}~\bibnamefont {González-Rodríguez}},\ }\bibfield  {title} {\bibinfo {title} {Understanding complex supramolecular landscapes: non-covalent macrocyclization equilibria examined by fluorescence resonance energy transfer},\ }\href {https://doi.org/10.1039/c8sc03229g} {\bibfield  {journal} {\bibinfo  {journal} {Chemical Science}\ }\textbf {\bibinfo {volume} {9}},\ \bibinfo {pages} {7809} (\bibinfo {year} {2018})}\BibitemShut {NoStop}%
\bibitem [{\citenamefont {Rajdev}\ and\ \citenamefont {Ghosh}(2019)}]{Raydev2019}%
  \BibitemOpen
  \bibfield  {author} {\bibinfo {author} {\bibfnamefont {P.}~\bibnamefont {Rajdev}}\ and\ \bibinfo {author} {\bibfnamefont {S.}~\bibnamefont {Ghosh}},\ }\bibfield  {title} {\bibinfo {title} {Fluorescence resonance energy transfer ({FRET}): A powerful tool for probing amphiphilic polymer aggregates and supramolecular polymers},\ }\href {https://pubs.acs.org/doi/abs/10.1021/acs.jpcb.8b09441} {\bibfield  {journal} {\bibinfo  {journal} {The Journal of Physical Chemistry B}\ }\textbf {\bibinfo {volume} {123}},\ \bibinfo {pages} {327} (\bibinfo {year} {2019})}\BibitemShut {NoStop}%
\bibitem [{\citenamefont {Banal}\ \emph {et~al.}(2017)\citenamefont {Banal}, \citenamefont {Zhang}, \citenamefont {Jones}, \citenamefont {Ghiggino},\ and\ \citenamefont {Wong}}]{Banal2017}%
  \BibitemOpen
  \bibfield  {author} {\bibinfo {author} {\bibfnamefont {J.~L.}\ \bibnamefont {Banal}}, \bibinfo {author} {\bibfnamefont {B.}~\bibnamefont {Zhang}}, \bibinfo {author} {\bibfnamefont {D.~J.}\ \bibnamefont {Jones}}, \bibinfo {author} {\bibfnamefont {K.~P.}\ \bibnamefont {Ghiggino}},\ and\ \bibinfo {author} {\bibfnamefont {W.~W.~H.}\ \bibnamefont {Wong}},\ }\bibfield  {title} {\bibinfo {title} {Emissive molecular aggregates and energy migration in luminescent solar concentrators},\ }\href {https://pubs.acs.org/doi/abs/10.1021/acs.accounts.6b00432} {\bibfield  {journal} {\bibinfo  {journal} {Accounts of Chemical Research}\ }\textbf {\bibinfo {volume} {50}},\ \bibinfo {pages} {49} (\bibinfo {year} {2017})}\BibitemShut {NoStop}%
\bibitem [{\citenamefont {Tummeltshammer}\ \emph {et~al.}(2017)\citenamefont {Tummeltshammer}, \citenamefont {Portnoi}, \citenamefont {Mitchell}, \citenamefont {Lee}, \citenamefont {Kenyon}, \citenamefont {Tabor},\ and\ \citenamefont {Papakonstantinou}}]{TUMMELTSHAMMER2017}%
  \BibitemOpen
  \bibfield  {author} {\bibinfo {author} {\bibfnamefont {C.}~\bibnamefont {Tummeltshammer}}, \bibinfo {author} {\bibfnamefont {M.}~\bibnamefont {Portnoi}}, \bibinfo {author} {\bibfnamefont {S.~A.}\ \bibnamefont {Mitchell}}, \bibinfo {author} {\bibfnamefont {A.-T.}\ \bibnamefont {Lee}}, \bibinfo {author} {\bibfnamefont {A.~J.}\ \bibnamefont {Kenyon}}, \bibinfo {author} {\bibfnamefont {A.~B.}\ \bibnamefont {Tabor}},\ and\ \bibinfo {author} {\bibfnamefont {I.}~\bibnamefont {Papakonstantinou}},\ }\bibfield  {title} {\bibinfo {title} {On the ability of {F}örster resonance energy transfer to enhance luminescent solar concentrator efficiency},\ }\href {https://doi.org/https://doi.org/10.1016/j.nanoen.2016.11.058} {\bibfield  {journal} {\bibinfo  {journal} {Nano Energy}\ }\textbf {\bibinfo {volume} {32}},\ \bibinfo {pages} {263} (\bibinfo {year} {2017})}\BibitemShut {NoStop}%
\bibitem [{\citenamefont {Zhang}\ \emph {et~al.}(2022)\citenamefont {Zhang}, \citenamefont {Lyu}, \citenamefont {Kelly},\ and\ \citenamefont {Evans}}]{Zhang2022}%
  \BibitemOpen
  \bibfield  {author} {\bibinfo {author} {\bibfnamefont {B.}~\bibnamefont {Zhang}}, \bibinfo {author} {\bibfnamefont {G.}~\bibnamefont {Lyu}}, \bibinfo {author} {\bibfnamefont {E.~A.}\ \bibnamefont {Kelly}},\ and\ \bibinfo {author} {\bibfnamefont {R.~C.}\ \bibnamefont {Evans}},\ }\bibfield  {title} {\bibinfo {title} {Förster resonance energy transfer in luminescent solar concentrators},\ }\href {https://doi.org/https://doi.org/10.1002/advs.202201160} {\bibfield  {journal} {\bibinfo  {journal} {Advanced Science}\ }\textbf {\bibinfo {volume} {9}},\ \bibinfo {pages} {2201160} (\bibinfo {year} {2022})}\BibitemShut {NoStop}%
\bibitem [{\citenamefont {Gorbenko}\ \emph {et~al.}(2017)\citenamefont {Gorbenko}, \citenamefont {Trusova},\ and\ \citenamefont {Molotkovsky}}]{Gorbenko2017}%
  \BibitemOpen
  \bibfield  {author} {\bibinfo {author} {\bibfnamefont {G.~P.}\ \bibnamefont {Gorbenko}}, \bibinfo {author} {\bibfnamefont {V.}~\bibnamefont {Trusova}},\ and\ \bibinfo {author} {\bibfnamefont {J.~G.}\ \bibnamefont {Molotkovsky}},\ }\bibfield  {title} {\bibinfo {title} {F\"{o}rster resonance energy transfer study of cytochrome c—lipid interactions},\ }\href {https://doi.org/10.1007/s10895-017-2176-1} {\bibfield  {journal} {\bibinfo  {journal} {Journal of Fluorescence}\ }\textbf {\bibinfo {volume} {28}},\ \bibinfo {pages} {79} (\bibinfo {year} {2017})}\BibitemShut {NoStop}%
\bibitem [{\citenamefont {Bartnik}\ \emph {et~al.}(2019)\citenamefont {Bartnik}, \citenamefont {Barth}, \citenamefont {Pilo-Pais}, \citenamefont {Crevenna}, \citenamefont {Liedl},\ and\ \citenamefont {Lamb}}]{Bartnik2019}%
  \BibitemOpen
  \bibfield  {author} {\bibinfo {author} {\bibfnamefont {K.}~\bibnamefont {Bartnik}}, \bibinfo {author} {\bibfnamefont {A.}~\bibnamefont {Barth}}, \bibinfo {author} {\bibfnamefont {M.}~\bibnamefont {Pilo-Pais}}, \bibinfo {author} {\bibfnamefont {A.~H.}\ \bibnamefont {Crevenna}}, \bibinfo {author} {\bibfnamefont {T.}~\bibnamefont {Liedl}},\ and\ \bibinfo {author} {\bibfnamefont {D.~C.}\ \bibnamefont {Lamb}},\ }\bibfield  {title} {\bibinfo {title} {A {DNA} origami platform for single-pair {F}\"{o}rster resonance energy transfer investigation of {DNA}–{DNA} interactions and ligation},\ }\href {https://doi.org/10.1021/jacs.9b09093} {\bibfield  {journal} {\bibinfo  {journal} {Journal of the American Chemical Society}\ }\textbf {\bibinfo {volume} {142}},\ \bibinfo {pages} {815} (\bibinfo {year} {2019})}\BibitemShut {NoStop}%
\bibitem [{\citenamefont {Fan}\ \emph {et~al.}(2023)\citenamefont {Fan}, \citenamefont {Wang}, \citenamefont {Yang}, \citenamefont {Zhong}, \citenamefont {Chen}, \citenamefont {Yu}, \citenamefont {Chen}, \citenamefont {Wu}, \citenamefont {Kuo}, \citenamefont {Lin},\ and\ \citenamefont {Chen}}]{Fan2023}%
  \BibitemOpen
  \bibfield  {author} {\bibinfo {author} {\bibfnamefont {X.}~\bibnamefont {Fan}}, \bibinfo {author} {\bibfnamefont {S.}~\bibnamefont {Wang}}, \bibinfo {author} {\bibfnamefont {X.}~\bibnamefont {Yang}}, \bibinfo {author} {\bibfnamefont {C.}~\bibnamefont {Zhong}}, \bibinfo {author} {\bibfnamefont {G.}~\bibnamefont {Chen}}, \bibinfo {author} {\bibfnamefont {C.}~\bibnamefont {Yu}}, \bibinfo {author} {\bibfnamefont {Y.}~\bibnamefont {Chen}}, \bibinfo {author} {\bibfnamefont {T.}~\bibnamefont {Wu}}, \bibinfo {author} {\bibfnamefont {H.}~\bibnamefont {Kuo}}, \bibinfo {author} {\bibfnamefont {Y.}~\bibnamefont {Lin}},\ and\ \bibinfo {author} {\bibfnamefont {Z.}~\bibnamefont {Chen}},\ }\bibfield  {title} {\bibinfo {title} {Brightened bicomponent perovskite nanocomposite based on {F}\"{o}rster resonance energy transfer for micro‐{LED} displays},\ }\href {http://dx.doi.org/10.1002/adma.202300834} {\bibfield  {journal} {\bibinfo  {journal} {Advanced Materials}\ }\textbf {\bibinfo {volume} {35}},\ \bibinfo {pages}
  {2300834} (\bibinfo {year} {2023})}\BibitemShut {NoStop}%
\bibitem [{\citenamefont {Li}\ \emph {et~al.}(2024)\citenamefont {Li}, \citenamefont {Qian}, \citenamefont {Liu},\ and\ \citenamefont {Qiu}}]{Li2024}%
  \BibitemOpen
  \bibfield  {author} {\bibinfo {author} {\bibfnamefont {Y.}~\bibnamefont {Li}}, \bibinfo {author} {\bibfnamefont {M.}~\bibnamefont {Qian}}, \bibinfo {author} {\bibfnamefont {Y.}~\bibnamefont {Liu}},\ and\ \bibinfo {author} {\bibfnamefont {X.}~\bibnamefont {Qiu}},\ }\bibfield  {title} {\bibinfo {title} {Approach: Sensitive detection of exosomal biomarkers by aptamer-mediated proximity ligation assay and time-resolved {F}\"{o}rster resonance energy transfer},\ }\href {http://dx.doi.org/10.3390/bios14050233} {\bibfield  {journal} {\bibinfo  {journal} {Biosensors}\ }\textbf {\bibinfo {volume} {14}},\ \bibinfo {pages} {233} (\bibinfo {year} {2024})}\BibitemShut {NoStop}%
\bibitem [{\citenamefont {Yablonovitch}(1994)}]{Yablonovitch1994}%
  \BibitemOpen
  \bibfield  {author} {\bibinfo {author} {\bibfnamefont {E.}~\bibnamefont {Yablonovitch}},\ }\bibfield  {title} {\bibinfo {title} {Photonic crystals},\ }\href {https://doi.org/10.1080/09500349414550261} {\bibfield  {journal} {\bibinfo  {journal} {Journal of Modern Optics}\ }\textbf {\bibinfo {volume} {41}},\ \bibinfo {pages} {173} (\bibinfo {year} {1994})}\BibitemShut {NoStop}%
\bibitem [{\citenamefont {Kort-Kamp}\ \emph {et~al.}(2014)\citenamefont {Kort-Kamp}, \citenamefont {Rosa}, \citenamefont {Pinheiro},\ and\ \citenamefont {Farina}}]{KortKamp2014}%
  \BibitemOpen
  \bibfield  {author} {\bibinfo {author} {\bibfnamefont {W.~J.~M.}\ \bibnamefont {Kort-Kamp}}, \bibinfo {author} {\bibfnamefont {F.~S.~S.}\ \bibnamefont {Rosa}}, \bibinfo {author} {\bibfnamefont {F.~A.}\ \bibnamefont {Pinheiro}},\ and\ \bibinfo {author} {\bibfnamefont {C.}~\bibnamefont {Farina}},\ }\bibfield  {title} {\bibinfo {title} {Molding the flow of light with a magnetic field: plasmonic cloaking and directional scattering},\ }\href {https://doi.org/10.1364/JOSAA.31.001969} {\bibfield  {journal} {\bibinfo  {journal} {JOSA A}\ }\textbf {\bibinfo {volume} {31}},\ \bibinfo {pages} {1969} (\bibinfo {year} {2014})}\BibitemShut {NoStop}%
\bibitem [{\citenamefont {Kort-Kamp}\ \emph {et~al.}(2015)\citenamefont {Kort-Kamp}, \citenamefont {Amorim}, \citenamefont {Bastos}, \citenamefont {Pinheiro}, \citenamefont {Rosa}, \citenamefont {Peres},\ and\ \citenamefont {Farina}}]{KortKamp2015}%
  \BibitemOpen
  \bibfield  {author} {\bibinfo {author} {\bibfnamefont {W.~J.~M.}\ \bibnamefont {Kort-Kamp}}, \bibinfo {author} {\bibfnamefont {B.}~\bibnamefont {Amorim}}, \bibinfo {author} {\bibfnamefont {G.}~\bibnamefont {Bastos}}, \bibinfo {author} {\bibfnamefont {F.~A.}\ \bibnamefont {Pinheiro}}, \bibinfo {author} {\bibfnamefont {F.~S.~S.}\ \bibnamefont {Rosa}}, \bibinfo {author} {\bibfnamefont {N.~M.~R.}\ \bibnamefont {Peres}},\ and\ \bibinfo {author} {\bibfnamefont {C.}~\bibnamefont {Farina}},\ }\bibfield  {title} {\bibinfo {title} {Active magneto-optical control of spontaneous emission in graphene},\ }\href {http://dx.doi.org/10.1103/PhysRevB.92.205415} {\bibfield  {journal} {\bibinfo  {journal} {Physical Review B}\ }\textbf {\bibinfo {volume} {92}},\ \bibinfo {pages} {205415} (\bibinfo {year} {2015})}\BibitemShut {NoStop}%
\bibitem [{\citenamefont {de~Melo~e Souza}\ \emph {et~al.}(2015)\citenamefont {de~Melo~e Souza}, \citenamefont {Kort-Kamp}, \citenamefont {Rosa},\ and\ \citenamefont {Farina}}]{Reinaldo2015}%
  \BibitemOpen
  \bibfield  {author} {\bibinfo {author} {\bibfnamefont {R.}~\bibnamefont {de~Melo~e Souza}}, \bibinfo {author} {\bibfnamefont {W.~J.~M.}\ \bibnamefont {Kort-Kamp}}, \bibinfo {author} {\bibfnamefont {F.~S.~S.}\ \bibnamefont {Rosa}},\ and\ \bibinfo {author} {\bibfnamefont {C.}~\bibnamefont {Farina}},\ }\bibfield  {title} {\bibinfo {title} {Influence of a surface in the nonretarded interaction between two atoms},\ }\href {https://doi.org/10.1103/PhysRevA.91.052708} {\bibfield  {journal} {\bibinfo  {journal} {Physical Review A}\ }\textbf {\bibinfo {volume} {91}},\ \bibinfo {pages} {052708} (\bibinfo {year} {2015})}\BibitemShut {NoStop}%
\bibitem [{\citenamefont {Szilard}\ \emph {et~al.}(2019)\citenamefont {Szilard}, \citenamefont {Kort-Kamp}, \citenamefont {Rosa}, \citenamefont {Pinheiro},\ and\ \citenamefont {Farina}}]{Szilard2019}%
  \BibitemOpen
  \bibfield  {author} {\bibinfo {author} {\bibfnamefont {D.}~\bibnamefont {Szilard}}, \bibinfo {author} {\bibfnamefont {W.}~\bibnamefont {Kort-Kamp}}, \bibinfo {author} {\bibfnamefont {F.}~\bibnamefont {Rosa}}, \bibinfo {author} {\bibfnamefont {F.}~\bibnamefont {Pinheiro}},\ and\ \bibinfo {author} {\bibfnamefont {C.}~\bibnamefont {Farina}},\ }\bibfield  {title} {\bibinfo {title} {Hysteresis in the spontaneous emission induced by {VO}$_2$ phase change},\ }\href {https://opg.optica.org/josab/abstract.cfm?uri=josab-36-4-c46} {\bibfield  {journal} {\bibinfo  {journal} {JOSA B}\ }\textbf {\bibinfo {volume} {36}},\ \bibinfo {pages} {C46} (\bibinfo {year} {2019})}\BibitemShut {NoStop}%
\bibitem [{\citenamefont {Chinh}(2020)}]{chinh2020}%
  \BibitemOpen
  \bibfield  {author} {\bibinfo {author} {\bibfnamefont {N.~D.}\ \bibnamefont {Chinh}},\ }\bibfield  {title} {\bibinfo {title} {Strong enhancement and inhibition of the interatomic van der {W}aals interaction inside a cylindrical waveguide},\ }\href {https://link.springer.com/article/10.1140/epjd/e2020-10024-9} {\bibfield  {journal} {\bibinfo  {journal} {The European Physical Journal D}\ }\textbf {\bibinfo {volume} {74}},\ \bibinfo {pages} {1} (\bibinfo {year} {2020})}\BibitemShut {NoStop}%
\bibitem [{\citenamefont {Casabone}\ \emph {et~al.}(2021)\citenamefont {Casabone}, \citenamefont {Deshmukh}, \citenamefont {Liu}, \citenamefont {Serrano}, \citenamefont {Ferrier}, \citenamefont {H\"{u}mmer}, \citenamefont {Goldner}, \citenamefont {Hunger},\ and\ \citenamefont {de~Riedmatten}}]{Casabone2021}%
  \BibitemOpen
  \bibfield  {author} {\bibinfo {author} {\bibfnamefont {B.}~\bibnamefont {Casabone}}, \bibinfo {author} {\bibfnamefont {C.}~\bibnamefont {Deshmukh}}, \bibinfo {author} {\bibfnamefont {S.}~\bibnamefont {Liu}}, \bibinfo {author} {\bibfnamefont {D.}~\bibnamefont {Serrano}}, \bibinfo {author} {\bibfnamefont {A.}~\bibnamefont {Ferrier}}, \bibinfo {author} {\bibfnamefont {T.}~\bibnamefont {H\"{u}mmer}}, \bibinfo {author} {\bibfnamefont {P.}~\bibnamefont {Goldner}}, \bibinfo {author} {\bibfnamefont {D.}~\bibnamefont {Hunger}},\ and\ \bibinfo {author} {\bibfnamefont {H.}~\bibnamefont {de~Riedmatten}},\ }\bibfield  {title} {\bibinfo {title} {Dynamic control of {P}urcell enhanced emission of erbium ions in nanoparticles},\ }\href {http://dx.doi.org/10.1038/s41467-021-23632-9} {\bibfield  {journal} {\bibinfo  {journal} {Nature Communications}\ }\textbf {\bibinfo {volume} {12}},\ \bibinfo {pages} {3570} (\bibinfo {year} {2021})}\BibitemShut {NoStop}%
\bibitem [{\citenamefont {Agarwal}(2024)}]{Agarwal2024}%
  \BibitemOpen
  \bibfield  {author} {\bibinfo {author} {\bibfnamefont {G.~S.}\ \bibnamefont {Agarwal}},\ }\bibfield  {title} {\bibinfo {title} {Control of the {P}urcell effect via unexcited atoms and exceptional points},\ }\href {http://dx.doi.org/10.1103/PhysRevResearch.6.L012050} {\bibfield  {journal} {\bibinfo  {journal} {Physical Review Research}\ }\textbf {\bibinfo {volume} {6}},\ \bibinfo {pages} {L012050} (\bibinfo {year} {2024})}\BibitemShut {NoStop}%
\bibitem [{\citenamefont {Barnes}(1998)}]{Barnes1998}%
  \BibitemOpen
  \bibfield  {author} {\bibinfo {author} {\bibfnamefont {W.~L.}\ \bibnamefont {Barnes}},\ }\bibfield  {title} {\bibinfo {title} {Fluorescence near interfaces: The role of photonic mode density},\ }\href {https://api.semanticscholar.org/CorpusID:120509536} {\bibfield  {journal} {\bibinfo  {journal} {Journal of Modern Optics}\ }\textbf {\bibinfo {volume} {45}},\ \bibinfo {pages} {661} (\bibinfo {year} {1998})}\BibitemShut {NoStop}%
\bibitem [{\citenamefont {Barnes}\ and\ \citenamefont {Andrew}(1999)}]{Barnes1999}%
  \BibitemOpen
  \bibfield  {author} {\bibinfo {author} {\bibfnamefont {W.}~\bibnamefont {Barnes}}\ and\ \bibinfo {author} {\bibfnamefont {P.}~\bibnamefont {Andrew}},\ }\bibfield  {title} {\bibinfo {title} {Quantum optics: Energy transfer under control},\ }\href {https://doi.org/10.1038/22875} {\bibfield  {journal} {\bibinfo  {journal} {Nature}\ }\textbf {\bibinfo {volume} {400}},\ \bibinfo {pages} {505} (\bibinfo {year} {1999})}\BibitemShut {NoStop}%
\bibitem [{\citenamefont {Abrantes}\ \emph {et~al.}(2020)\citenamefont {Abrantes}, \citenamefont {Szilard}, \citenamefont {Rosa},\ and\ \citenamefont {Farina}}]{Abrantes2020}%
  \BibitemOpen
  \bibfield  {author} {\bibinfo {author} {\bibfnamefont {P.~P.}\ \bibnamefont {Abrantes}}, \bibinfo {author} {\bibfnamefont {D.}~\bibnamefont {Szilard}}, \bibinfo {author} {\bibfnamefont {F.~S.~S.}\ \bibnamefont {Rosa}},\ and\ \bibinfo {author} {\bibfnamefont {C.}~\bibnamefont {Farina}},\ }\bibfield  {title} {\bibinfo {title} {Resonance energy transfer at percolation transition},\ }\href {https://doi.org/10.1142/s0217732320400222} {\bibfield  {journal} {\bibinfo  {journal} {Modern Physics Letters A}\ }\textbf {\bibinfo {volume} {35}},\ \bibinfo {pages} {2040022} (\bibinfo {year} {2020})}\BibitemShut {NoStop}%
\bibitem [{\citenamefont {Abrantes}\ \emph {et~al.}(2021{\natexlab{a}})\citenamefont {Abrantes}, \citenamefont {Bastos}, \citenamefont {Szilard}, \citenamefont {Farina},\ and\ \citenamefont {Rosa}}]{patricia2021}%
  \BibitemOpen
  \bibfield  {author} {\bibinfo {author} {\bibfnamefont {P.~P.}\ \bibnamefont {Abrantes}}, \bibinfo {author} {\bibfnamefont {G.}~\bibnamefont {Bastos}}, \bibinfo {author} {\bibfnamefont {D.}~\bibnamefont {Szilard}}, \bibinfo {author} {\bibfnamefont {C.}~\bibnamefont {Farina}},\ and\ \bibinfo {author} {\bibfnamefont {F.~S.~S.}\ \bibnamefont {Rosa}},\ }\bibfield  {title} {\bibinfo {title} {Tuning resonance energy transfer with magneto-optical properties of graphene},\ }\href {https://doi.org/10.1103/PhysRevB.103.174421} {\bibfield  {journal} {\bibinfo  {journal} {Physical Review B}\ }\textbf {\bibinfo {volume} {103}},\ \bibinfo {pages} {174421} (\bibinfo {year} {2021}{\natexlab{a}})}\BibitemShut {NoStop}%
\bibitem [{\citenamefont {Nayem}\ \emph {et~al.}(2023)\citenamefont {Nayem}, \citenamefont {Sikder},\ and\ \citenamefont {Uddin}}]{Nayem2023}%
  \BibitemOpen
  \bibfield  {author} {\bibinfo {author} {\bibfnamefont {S.~H.}\ \bibnamefont {Nayem}}, \bibinfo {author} {\bibfnamefont {B.}~\bibnamefont {Sikder}},\ and\ \bibinfo {author} {\bibfnamefont {S.~Z.}\ \bibnamefont {Uddin}},\ }\bibfield  {title} {\bibinfo {title} {Anisotropic energy transfer near multi-layer black phosphorus},\ }\href {https://doi.org/10.1088/2053-1583/acf052} {\bibfield  {journal} {\bibinfo  {journal} {2D Materials}\ }\textbf {\bibinfo {volume} {10}},\ \bibinfo {pages} {045022} (\bibinfo {year} {2023})}\BibitemShut {NoStop}%
\bibitem [{\citenamefont {Lezhennikova}\ \emph {et~al.}(2023)\citenamefont {Lezhennikova}, \citenamefont {Rustomji}, \citenamefont {Kuhlmey}, \citenamefont {Antonakakis}, \citenamefont {Jomin}, \citenamefont {Glybovski}, \citenamefont {de~Sterke}, \citenamefont {Wenger}, \citenamefont {Abdeddaim},\ and\ \citenamefont {Enoch}}]{Lezhennikova2023}%
  \BibitemOpen
  \bibfield  {author} {\bibinfo {author} {\bibfnamefont {K.}~\bibnamefont {Lezhennikova}}, \bibinfo {author} {\bibfnamefont {K.}~\bibnamefont {Rustomji}}, \bibinfo {author} {\bibfnamefont {B.~T.}\ \bibnamefont {Kuhlmey}}, \bibinfo {author} {\bibfnamefont {T.}~\bibnamefont {Antonakakis}}, \bibinfo {author} {\bibfnamefont {P.}~\bibnamefont {Jomin}}, \bibinfo {author} {\bibfnamefont {S.}~\bibnamefont {Glybovski}}, \bibinfo {author} {\bibfnamefont {C.~M.}\ \bibnamefont {de~Sterke}}, \bibinfo {author} {\bibfnamefont {J.}~\bibnamefont {Wenger}}, \bibinfo {author} {\bibfnamefont {R.}~\bibnamefont {Abdeddaim}},\ and\ \bibinfo {author} {\bibfnamefont {S.}~\bibnamefont {Enoch}},\ }\bibfield  {title} {\bibinfo {title} {Experimental evidence of {F}\"{o}rster energy transfer enhancement in the near field through engineered metamaterial surface waves},\ }\href {http://dx.doi.org/10.1038/s42005-023-01347-1} {\bibfield  {journal} {\bibinfo  {journal} {Communications Physics}\ }\textbf {\bibinfo {volume} {6}},\ \bibinfo
  {pages} {229} (\bibinfo {year} {2023})}\BibitemShut {NoStop}%
\bibitem [{\citenamefont {Beutler}\ \emph {et~al.}(2024)\citenamefont {Beutler}, \citenamefont {Camden},\ and\ \citenamefont {Masiello}}]{Beutler2024}%
  \BibitemOpen
  \bibfield  {author} {\bibinfo {author} {\bibfnamefont {E.~K.}\ \bibnamefont {Beutler}}, \bibinfo {author} {\bibfnamefont {J.~P.}\ \bibnamefont {Camden}},\ and\ \bibinfo {author} {\bibfnamefont {D.~J.}\ \bibnamefont {Masiello}},\ }\bibfield  {title} {\bibinfo {title} {Resonance energy transfer in the presence of a spherical cavity and dispersive substrate},\ }\href {https://doi.org/10.1103/PhysRevB.110.155435} {\bibfield  {journal} {\bibinfo  {journal} {Physical Review B}\ }\textbf {\bibinfo {volume} {110}},\ \bibinfo {pages} {155435} (\bibinfo {year} {2024})}\BibitemShut {NoStop}%
\bibitem [{\citenamefont {Reserbat-Plantey}\ \emph {et~al.}(2021)\citenamefont {Reserbat-Plantey}, \citenamefont {Epstein}, \citenamefont {Torre}, \citenamefont {Costa}, \citenamefont {Gonçalves}, \citenamefont {Mortensen}, \citenamefont {Polini}, \citenamefont {Song}, \citenamefont {Peres},\ and\ \citenamefont {Koppens}}]{Plantey2021}%
  \BibitemOpen
  \bibfield  {author} {\bibinfo {author} {\bibfnamefont {A.}~\bibnamefont {Reserbat-Plantey}}, \bibinfo {author} {\bibfnamefont {I.}~\bibnamefont {Epstein}}, \bibinfo {author} {\bibfnamefont {I.}~\bibnamefont {Torre}}, \bibinfo {author} {\bibfnamefont {A.~T.}\ \bibnamefont {Costa}}, \bibinfo {author} {\bibfnamefont {P.~A.~D.}\ \bibnamefont {Gonçalves}}, \bibinfo {author} {\bibfnamefont {N.~A.}\ \bibnamefont {Mortensen}}, \bibinfo {author} {\bibfnamefont {M.}~\bibnamefont {Polini}}, \bibinfo {author} {\bibfnamefont {J.~C.~W.}\ \bibnamefont {Song}}, \bibinfo {author} {\bibfnamefont {N.~M.~R.}\ \bibnamefont {Peres}},\ and\ \bibinfo {author} {\bibfnamefont {F.~H.~L.}\ \bibnamefont {Koppens}},\ }\bibfield  {title} {\bibinfo {title} {Quantum nanophotonics in two-dimensional materials},\ }\href {https://doi.org/10.1021/acsphotonics.0c01224} {\bibfield  {journal} {\bibinfo  {journal} {ACS Photonics}\ }\textbf {\bibinfo {volume} {8}},\ \bibinfo {pages} {85} (\bibinfo {year} {2021})}\BibitemShut {NoStop}%
\bibitem [{\citenamefont {Meng}\ \emph {et~al.}(2023)\citenamefont {Meng}, \citenamefont {Zhong}, \citenamefont {Xu}, \citenamefont {He}, \citenamefont {Kim}, \citenamefont {Han}, \citenamefont {Kim}, \citenamefont {Park}, \citenamefont {Shen}, \citenamefont {Gong}, \citenamefont {Xiao},\ and\ \citenamefont {Bae}}]{Meng2023}%
  \BibitemOpen
  \bibfield  {author} {\bibinfo {author} {\bibfnamefont {Y.}~\bibnamefont {Meng}}, \bibinfo {author} {\bibfnamefont {H.}~\bibnamefont {Zhong}}, \bibinfo {author} {\bibfnamefont {Z.}~\bibnamefont {Xu}}, \bibinfo {author} {\bibfnamefont {T.}~\bibnamefont {He}}, \bibinfo {author} {\bibfnamefont {J.~S.}\ \bibnamefont {Kim}}, \bibinfo {author} {\bibfnamefont {S.}~\bibnamefont {Han}}, \bibinfo {author} {\bibfnamefont {S.}~\bibnamefont {Kim}}, \bibinfo {author} {\bibfnamefont {S.}~\bibnamefont {Park}}, \bibinfo {author} {\bibfnamefont {Y.}~\bibnamefont {Shen}}, \bibinfo {author} {\bibfnamefont {M.}~\bibnamefont {Gong}}, \bibinfo {author} {\bibfnamefont {Q.}~\bibnamefont {Xiao}},\ and\ \bibinfo {author} {\bibfnamefont {S.-H.}\ \bibnamefont {Bae}},\ }\bibfield  {title} {\bibinfo {title} {Functionalizing nanophotonic structures with 2{D} van der {W}aals materials},\ }\href {https://doi.org/10.1039/d3nh00246b} {\bibfield  {journal} {\bibinfo  {journal} {Nanoscale Horizons}\ }\textbf {\bibinfo {volume} {8}},\ \bibinfo
  {pages} {1345} (\bibinfo {year} {2023})}\BibitemShut {NoStop}%
\bibitem [{\citenamefont {Banishev}\ \emph {et~al.}(2013)\citenamefont {Banishev}, \citenamefont {Wen}, \citenamefont {Xu}, \citenamefont {Kawakami}, \citenamefont {Klimchitskaya}, \citenamefont {Mostepanenko},\ and\ \citenamefont {Mohideen}}]{Banishev2013}%
  \BibitemOpen
  \bibfield  {author} {\bibinfo {author} {\bibfnamefont {A.~A.}\ \bibnamefont {Banishev}}, \bibinfo {author} {\bibfnamefont {H.}~\bibnamefont {Wen}}, \bibinfo {author} {\bibfnamefont {J.}~\bibnamefont {Xu}}, \bibinfo {author} {\bibfnamefont {R.~K.}\ \bibnamefont {Kawakami}}, \bibinfo {author} {\bibfnamefont {G.~L.}\ \bibnamefont {Klimchitskaya}}, \bibinfo {author} {\bibfnamefont {V.~M.}\ \bibnamefont {Mostepanenko}},\ and\ \bibinfo {author} {\bibfnamefont {U.}~\bibnamefont {Mohideen}},\ }\bibfield  {title} {\bibinfo {title} {Measuring the {C}asimir force gradient from graphene on a {S}i{O}$_{2}$ substrate},\ }\href {https://doi.org/10.1103/PhysRevB.87.205433} {\bibfield  {journal} {\bibinfo  {journal} {Physical Review B}\ }\textbf {\bibinfo {volume} {87}},\ \bibinfo {pages} {205433} (\bibinfo {year} {2013})}\BibitemShut {NoStop}%
\bibitem [{\citenamefont {Cysne}\ \emph {et~al.}(2014)\citenamefont {Cysne}, \citenamefont {Kort-Kamp}, \citenamefont {Oliver}, \citenamefont {Pinheiro}, \citenamefont {Rosa},\ and\ \citenamefont {Farina}}]{Cysne2014}%
  \BibitemOpen
  \bibfield  {author} {\bibinfo {author} {\bibfnamefont {T.}~\bibnamefont {Cysne}}, \bibinfo {author} {\bibfnamefont {W.~J.~M.}\ \bibnamefont {Kort-Kamp}}, \bibinfo {author} {\bibfnamefont {D.}~\bibnamefont {Oliver}}, \bibinfo {author} {\bibfnamefont {F.~A.}\ \bibnamefont {Pinheiro}}, \bibinfo {author} {\bibfnamefont {F.~S.~S.}\ \bibnamefont {Rosa}},\ and\ \bibinfo {author} {\bibfnamefont {C.}~\bibnamefont {Farina}},\ }\bibfield  {title} {\bibinfo {title} {Tuning the {C}asimir-{P}older interaction via magneto-optical effects in graphene},\ }\href {https://doi.org/10.1103/PhysRevA.90.052511} {\bibfield  {journal} {\bibinfo  {journal} {Physical Review A}\ }\textbf {\bibinfo {volume} {90}},\ \bibinfo {pages} {052511} (\bibinfo {year} {2014})}\BibitemShut {NoStop}%
\bibitem [{\citenamefont {Woods}\ \emph {et~al.}(2016)\citenamefont {Woods}, \citenamefont {Dalvit}, \citenamefont {Tkatchenko}, \citenamefont {Rodriguez-Lopez}, \citenamefont {Rodriguez},\ and\ \citenamefont {Podgornik}}]{Woods2016}%
  \BibitemOpen
  \bibfield  {author} {\bibinfo {author} {\bibfnamefont {L.~M.}\ \bibnamefont {Woods}}, \bibinfo {author} {\bibfnamefont {D.~A.~R.}\ \bibnamefont {Dalvit}}, \bibinfo {author} {\bibfnamefont {A.}~\bibnamefont {Tkatchenko}}, \bibinfo {author} {\bibfnamefont {P.}~\bibnamefont {Rodriguez-Lopez}}, \bibinfo {author} {\bibfnamefont {A.~W.}\ \bibnamefont {Rodriguez}},\ and\ \bibinfo {author} {\bibfnamefont {R.}~\bibnamefont {Podgornik}},\ }\bibfield  {title} {\bibinfo {title} {Materials perspective on {C}asimir and van der {W}aals interactions},\ }\href {https://doi.org/10.1103/RevModPhys.88.045003} {\bibfield  {journal} {\bibinfo  {journal} {Reviews of Modern Physics}\ }\textbf {\bibinfo {volume} {88}},\ \bibinfo {pages} {045003} (\bibinfo {year} {2016})}\BibitemShut {NoStop}%
\bibitem [{\citenamefont {Rodriguez-Lopez}\ \emph {et~al.}(2017)\citenamefont {Rodriguez-Lopez}, \citenamefont {Kort-Kamp}, \citenamefont {Dalvit},\ and\ \citenamefont {Woods}}]{RodriguezLopez2017}%
  \BibitemOpen
  \bibfield  {author} {\bibinfo {author} {\bibfnamefont {P.}~\bibnamefont {Rodriguez-Lopez}}, \bibinfo {author} {\bibfnamefont {W.~J.~M.}\ \bibnamefont {Kort-Kamp}}, \bibinfo {author} {\bibfnamefont {D.~A.~R.}\ \bibnamefont {Dalvit}},\ and\ \bibinfo {author} {\bibfnamefont {L.~M.}\ \bibnamefont {Woods}},\ }\bibfield  {title} {\bibinfo {title} {Casimir force phase transitions in the graphene family},\ }\href {http://dx.doi.org/10.1038/ncomms14699} {\bibfield  {journal} {\bibinfo  {journal} {Nature Communications}\ }\textbf {\bibinfo {volume} {8}},\ \bibinfo {pages} {14699} (\bibinfo {year} {2017})}\BibitemShut {NoStop}%
\bibitem [{\citenamefont {Silvestre}\ \emph {et~al.}(2019)\citenamefont {Silvestre}, \citenamefont {Cysne}, \citenamefont {Szilard}, \citenamefont {Pinheiro},\ and\ \citenamefont {Farina}}]{Marcus2019}%
  \BibitemOpen
  \bibfield  {author} {\bibinfo {author} {\bibfnamefont {M.}~\bibnamefont {Silvestre}}, \bibinfo {author} {\bibfnamefont {T.~P.}\ \bibnamefont {Cysne}}, \bibinfo {author} {\bibfnamefont {D.}~\bibnamefont {Szilard}}, \bibinfo {author} {\bibfnamefont {F.~A.}\ \bibnamefont {Pinheiro}},\ and\ \bibinfo {author} {\bibfnamefont {C.}~\bibnamefont {Farina}},\ }\bibfield  {title} {\bibinfo {title} {Tuning quantum reflection in graphene with an external magnetic field},\ }\href {https://doi.org/10.1103/PhysRevA.100.033605} {\bibfield  {journal} {\bibinfo  {journal} {Physical Review A}\ }\textbf {\bibinfo {volume} {100}},\ \bibinfo {pages} {033605} (\bibinfo {year} {2019})}\BibitemShut {NoStop}%
\bibitem [{\citenamefont {Liu}\ \emph {et~al.}(2021)\citenamefont {Liu}, \citenamefont {Zhang}, \citenamefont {Klimchitskaya}, \citenamefont {Mostepanenko},\ and\ \citenamefont {Mohideen}}]{Liu2021}%
  \BibitemOpen
  \bibfield  {author} {\bibinfo {author} {\bibfnamefont {M.}~\bibnamefont {Liu}}, \bibinfo {author} {\bibfnamefont {Y.}~\bibnamefont {Zhang}}, \bibinfo {author} {\bibfnamefont {G.~L.}\ \bibnamefont {Klimchitskaya}}, \bibinfo {author} {\bibfnamefont {V.~M.}\ \bibnamefont {Mostepanenko}},\ and\ \bibinfo {author} {\bibfnamefont {U.}~\bibnamefont {Mohideen}},\ }\bibfield  {title} {\bibinfo {title} {Demonstration of an unusual thermal effect in the {C}asimir force from graphene},\ }\href {https://doi.org/10.1103/PhysRevLett.126.206802} {\bibfield  {journal} {\bibinfo  {journal} {Physical Review Letters}\ }\textbf {\bibinfo {volume} {126}},\ \bibinfo {pages} {206802} (\bibinfo {year} {2021})}\BibitemShut {NoStop}%
\bibitem [{\citenamefont {Abrantes}\ \emph {et~al.}(2021{\natexlab{b}})\citenamefont {Abrantes}, \citenamefont {Cysne}, \citenamefont {Szilard}, \citenamefont {Rosa}, \citenamefont {Pinheiro},\ and\ \citenamefont {Farina}}]{Abrantes2021b}%
  \BibitemOpen
  \bibfield  {author} {\bibinfo {author} {\bibfnamefont {P.~P.}\ \bibnamefont {Abrantes}}, \bibinfo {author} {\bibfnamefont {T.~P.}\ \bibnamefont {Cysne}}, \bibinfo {author} {\bibfnamefont {D.}~\bibnamefont {Szilard}}, \bibinfo {author} {\bibfnamefont {F.~S.~S.}\ \bibnamefont {Rosa}}, \bibinfo {author} {\bibfnamefont {F.~A.}\ \bibnamefont {Pinheiro}},\ and\ \bibinfo {author} {\bibfnamefont {C.}~\bibnamefont {Farina}},\ }\bibfield  {title} {\bibinfo {title} {Probing topological phase transitions via quantum reflection in the graphene family materials},\ }\href {https://doi.org/10.1103/PhysRevB.104.075409} {\bibfield  {journal} {\bibinfo  {journal} {Physical Review B}\ }\textbf {\bibinfo {volume} {104}},\ \bibinfo {pages} {075409} (\bibinfo {year} {2021}{\natexlab{b}})}\BibitemShut {NoStop}%
\bibitem [{\citenamefont {Gaudreau}\ \emph {et~al.}(2013)\citenamefont {Gaudreau}, \citenamefont {Tielrooij}, \citenamefont {Prawiroatmodjo}, \citenamefont {Osmond}, \citenamefont {de~Abajo},\ and\ \citenamefont {Koppens}}]{Gaudreau2013}%
  \BibitemOpen
  \bibfield  {author} {\bibinfo {author} {\bibfnamefont {L.}~\bibnamefont {Gaudreau}}, \bibinfo {author} {\bibfnamefont {K.~J.}\ \bibnamefont {Tielrooij}}, \bibinfo {author} {\bibfnamefont {G.~E. D.~K.}\ \bibnamefont {Prawiroatmodjo}}, \bibinfo {author} {\bibfnamefont {J.}~\bibnamefont {Osmond}}, \bibinfo {author} {\bibfnamefont {F.~J.~G.}\ \bibnamefont {de~Abajo}},\ and\ \bibinfo {author} {\bibfnamefont {F.~H.~L.}\ \bibnamefont {Koppens}},\ }\bibfield  {title} {\bibinfo {title} {Universal distance-scaling of nonradiative energy transfer to graphene},\ }\href {https://doi.org/10.1021/nl400176b} {\bibfield  {journal} {\bibinfo  {journal} {Nano Letters}\ }\textbf {\bibinfo {volume} {13}},\ \bibinfo {pages} {2030} (\bibinfo {year} {2013})}\BibitemShut {NoStop}%
\bibitem [{\citenamefont {Raja}\ \emph {et~al.}(2016)\citenamefont {Raja}, \citenamefont {Montoya-Castillo}, \citenamefont {Zultak}, \citenamefont {Zhang}, \citenamefont {Ye}, \citenamefont {Roquelet}, \citenamefont {Chenet}, \citenamefont {van~der Zande}, \citenamefont {Huang}, \citenamefont {Jockusch}, \citenamefont {Hone}, \citenamefont {Reichman}, \citenamefont {Brus},\ and\ \citenamefont {Heinz}}]{Raja2016}%
  \BibitemOpen
  \bibfield  {author} {\bibinfo {author} {\bibfnamefont {A.}~\bibnamefont {Raja}}, \bibinfo {author} {\bibfnamefont {A.}~\bibnamefont {Montoya-Castillo}}, \bibinfo {author} {\bibfnamefont {J.}~\bibnamefont {Zultak}}, \bibinfo {author} {\bibfnamefont {X.-X.}\ \bibnamefont {Zhang}}, \bibinfo {author} {\bibfnamefont {Z.}~\bibnamefont {Ye}}, \bibinfo {author} {\bibfnamefont {C.}~\bibnamefont {Roquelet}}, \bibinfo {author} {\bibfnamefont {D.~A.}\ \bibnamefont {Chenet}}, \bibinfo {author} {\bibfnamefont {A.~M.}\ \bibnamefont {van~der Zande}}, \bibinfo {author} {\bibfnamefont {P.}~\bibnamefont {Huang}}, \bibinfo {author} {\bibfnamefont {S.}~\bibnamefont {Jockusch}}, \bibinfo {author} {\bibfnamefont {J.}~\bibnamefont {Hone}}, \bibinfo {author} {\bibfnamefont {D.~R.}\ \bibnamefont {Reichman}}, \bibinfo {author} {\bibfnamefont {L.~E.}\ \bibnamefont {Brus}},\ and\ \bibinfo {author} {\bibfnamefont {T.~F.}\ \bibnamefont {Heinz}},\ }\bibfield  {title} {\bibinfo {title} {Energy transfer from quantum dots to graphene and
  {M}o{S}$_2$: The role of absorption and screening in two-dimensional materials},\ }\href {https://doi.org/10.1021/acs.nanolett.5b05012} {\bibfield  {journal} {\bibinfo  {journal} {Nano Letters}\ }\textbf {\bibinfo {volume} {16}},\ \bibinfo {pages} {2328} (\bibinfo {year} {2016})}\BibitemShut {NoStop}%
\bibitem [{\citenamefont {Abrantes}\ \emph {et~al.}(2023)\citenamefont {Abrantes}, \citenamefont {Kort-Kamp}, \citenamefont {Rosa}, \citenamefont {Farina}, \citenamefont {Pinheiro},\ and\ \citenamefont {Cysne}}]{Patricia2024}%
  \BibitemOpen
  \bibfield  {author} {\bibinfo {author} {\bibfnamefont {P.~P.}\ \bibnamefont {Abrantes}}, \bibinfo {author} {\bibfnamefont {W.~J.~M.}\ \bibnamefont {Kort-Kamp}}, \bibinfo {author} {\bibfnamefont {F.~S.~S.}\ \bibnamefont {Rosa}}, \bibinfo {author} {\bibfnamefont {C.}~\bibnamefont {Farina}}, \bibinfo {author} {\bibfnamefont {F.~A.}\ \bibnamefont {Pinheiro}},\ and\ \bibinfo {author} {\bibfnamefont {T.~P.}\ \bibnamefont {Cysne}},\ }\bibfield  {title} {\bibinfo {title} {Controlling electric and magnetic {P}urcell effects in phosphorene via strain engineering},\ }\href {https://doi.org/10.1103/PhysRevB.108.155427} {\bibfield  {journal} {\bibinfo  {journal} {Physical Review B}\ }\textbf {\bibinfo {volume} {108}},\ \bibinfo {pages} {155427} (\bibinfo {year} {2023})}\BibitemShut {NoStop}%
\bibitem [{\citenamefont {Cavicchi}\ \emph {et~al.}(2024)\citenamefont {Cavicchi}, \citenamefont {Peralta}, \citenamefont {Álvaro Moreno}, \citenamefont {Vergniory}, \citenamefont {Jarillo-Herrero}, \citenamefont {Felser}, \citenamefont {Rocca}, \citenamefont {Koppens},\ and\ \citenamefont {Polini}}]{Cavicchi2024}%
  \BibitemOpen
  \bibfield  {author} {\bibinfo {author} {\bibfnamefont {L.}~\bibnamefont {Cavicchi}}, \bibinfo {author} {\bibfnamefont {M.}~\bibnamefont {Peralta}}, \bibinfo {author} {\bibnamefont {Álvaro Moreno}}, \bibinfo {author} {\bibfnamefont {M.}~\bibnamefont {Vergniory}}, \bibinfo {author} {\bibfnamefont {P.}~\bibnamefont {Jarillo-Herrero}}, \bibinfo {author} {\bibfnamefont {C.}~\bibnamefont {Felser}}, \bibinfo {author} {\bibfnamefont {G.~C.~L.}\ \bibnamefont {Rocca}}, \bibinfo {author} {\bibfnamefont {F.~H.~L.}\ \bibnamefont {Koppens}},\ and\ \bibinfo {author} {\bibfnamefont {M.}~\bibnamefont {Polini}},\ }\href@noop {} {\bibinfo {title} {Recognizing molecular chirality via twisted 2{D} materials}} (\bibinfo {year} {2024}),\ \Eprint {https://arxiv.org/abs/2409.05839} {arXiv:2409.05839 [cond-mat.mes-hall]} \BibitemShut {NoStop}%
\bibitem [{\citenamefont {Zhao}\ \emph {et~al.}(2017)\citenamefont {Zhao}, \citenamefont {Guizal}, \citenamefont {Zhang}, \citenamefont {Fan},\ and\ \citenamefont {Antezza}}]{Zhao2017}%
  \BibitemOpen
  \bibfield  {author} {\bibinfo {author} {\bibfnamefont {B.}~\bibnamefont {Zhao}}, \bibinfo {author} {\bibfnamefont {B.}~\bibnamefont {Guizal}}, \bibinfo {author} {\bibfnamefont {Z.~M.}\ \bibnamefont {Zhang}}, \bibinfo {author} {\bibfnamefont {S.}~\bibnamefont {Fan}},\ and\ \bibinfo {author} {\bibfnamefont {M.}~\bibnamefont {Antezza}},\ }\bibfield  {title} {\bibinfo {title} {Near-field heat transfer between graphene/h{BN} multilayers},\ }\href {https://doi.org/10.1103/PhysRevB.95.245437} {\bibfield  {journal} {\bibinfo  {journal} {Physical Review B}\ }\textbf {\bibinfo {volume} {95}},\ \bibinfo {pages} {245437} (\bibinfo {year} {2017})}\BibitemShut {NoStop}%
\bibitem [{\citenamefont {Ge}\ \emph {et~al.}(2019)\citenamefont {Ge}, \citenamefont {Gong}, \citenamefont {Cang}, \citenamefont {Luo}, \citenamefont {Shi},\ and\ \citenamefont {Wu}}]{Ge2019}%
  \BibitemOpen
  \bibfield  {author} {\bibinfo {author} {\bibfnamefont {L.}~\bibnamefont {Ge}}, \bibinfo {author} {\bibfnamefont {K.}~\bibnamefont {Gong}}, \bibinfo {author} {\bibfnamefont {Y.}~\bibnamefont {Cang}}, \bibinfo {author} {\bibfnamefont {Y.}~\bibnamefont {Luo}}, \bibinfo {author} {\bibfnamefont {X.}~\bibnamefont {Shi}},\ and\ \bibinfo {author} {\bibfnamefont {Y.}~\bibnamefont {Wu}},\ }\bibfield  {title} {\bibinfo {title} {Magnetically tunable multiband near-field radiative heat transfer between two graphene sheets},\ }\href {https://doi.org/10.1103/PhysRevB.100.035414} {\bibfield  {journal} {\bibinfo  {journal} {Physical Review B}\ }\textbf {\bibinfo {volume} {100}},\ \bibinfo {pages} {035414} (\bibinfo {year} {2019})}\BibitemShut {NoStop}%
\bibitem [{\citenamefont {Wu}\ \emph {et~al.}(2019)\citenamefont {Wu}, \citenamefont {Huang}, \citenamefont {Cui},\ and\ \citenamefont {Zhu}}]{Wu2019}%
  \BibitemOpen
  \bibfield  {author} {\bibinfo {author} {\bibfnamefont {H.}~\bibnamefont {Wu}}, \bibinfo {author} {\bibfnamefont {Y.}~\bibnamefont {Huang}}, \bibinfo {author} {\bibfnamefont {L.}~\bibnamefont {Cui}},\ and\ \bibinfo {author} {\bibfnamefont {K.}~\bibnamefont {Zhu}},\ }\bibfield  {title} {\bibinfo {title} {Active magneto-optical control of near-field radiative heat transfer between graphene sheets},\ }\href {https://doi.org/10.1103/PhysRevApplied.11.054020} {\bibfield  {journal} {\bibinfo  {journal} {Physical Review Applied}\ }\textbf {\bibinfo {volume} {11}},\ \bibinfo {pages} {054020} (\bibinfo {year} {2019})}\BibitemShut {NoStop}%
\bibitem [{\citenamefont {Iqbal}\ \emph {et~al.}(2023)\citenamefont {Iqbal}, \citenamefont {Zhang}, \citenamefont {Wang}, \citenamefont {Fang}, \citenamefont {Hu}, \citenamefont {Dang}, \citenamefont {Zhang}, \citenamefont {Jin}, \citenamefont {Xu}, \citenamefont {Ju},\ and\ \citenamefont {Ma}}]{Iqbal2023}%
  \BibitemOpen
  \bibfield  {author} {\bibinfo {author} {\bibfnamefont {N.}~\bibnamefont {Iqbal}}, \bibinfo {author} {\bibfnamefont {S.}~\bibnamefont {Zhang}}, \bibinfo {author} {\bibfnamefont {S.}~\bibnamefont {Wang}}, \bibinfo {author} {\bibfnamefont {Z.}~\bibnamefont {Fang}}, \bibinfo {author} {\bibfnamefont {Y.}~\bibnamefont {Hu}}, \bibinfo {author} {\bibfnamefont {Y.}~\bibnamefont {Dang}}, \bibinfo {author} {\bibfnamefont {M.}~\bibnamefont {Zhang}}, \bibinfo {author} {\bibfnamefont {Y.}~\bibnamefont {Jin}}, \bibinfo {author} {\bibfnamefont {J.}~\bibnamefont {Xu}}, \bibinfo {author} {\bibfnamefont {B.}~\bibnamefont {Ju}},\ and\ \bibinfo {author} {\bibfnamefont {Y.}~\bibnamefont {Ma}},\ }\bibfield  {title} {\bibinfo {title} {Measuring near-field radiative heat transfer in a graphene-$\mathrm{Si}\mathrm{C}$ heterostructure},\ }\href {https://doi.org/10.1103/PhysRevApplied.19.024019} {\bibfield  {journal} {\bibinfo  {journal} {Physical Review Applied}\ }\textbf {\bibinfo {volume} {19}},\ \bibinfo {pages} {024019} (\bibinfo
  {year} {2023})}\BibitemShut {NoStop}%
\bibitem [{\citenamefont {Li}\ \emph {et~al.}(2014)\citenamefont {Li}, \citenamefont {Yu}, \citenamefont {Ye}, \citenamefont {Ge}, \citenamefont {Ou}, \citenamefont {Wu}, \citenamefont {Feng}, \citenamefont {Chen},\ and\ \citenamefont {Zhang}}]{Li2014}%
  \BibitemOpen
  \bibfield  {author} {\bibinfo {author} {\bibfnamefont {L.}~\bibnamefont {Li}}, \bibinfo {author} {\bibfnamefont {Y.}~\bibnamefont {Yu}}, \bibinfo {author} {\bibfnamefont {G.~J.}\ \bibnamefont {Ye}}, \bibinfo {author} {\bibfnamefont {Q.}~\bibnamefont {Ge}}, \bibinfo {author} {\bibfnamefont {X.}~\bibnamefont {Ou}}, \bibinfo {author} {\bibfnamefont {H.}~\bibnamefont {Wu}}, \bibinfo {author} {\bibfnamefont {D.}~\bibnamefont {Feng}}, \bibinfo {author} {\bibfnamefont {X.~H.}\ \bibnamefont {Chen}},\ and\ \bibinfo {author} {\bibfnamefont {Y.}~\bibnamefont {Zhang}},\ }\bibfield  {title} {\bibinfo {title} {Black phosphorus field-effect transistors},\ }\href {http://dx.doi.org/10.1038/nnano.2014.35} {\bibfield  {journal} {\bibinfo  {journal} {Nature Nanotechnology}\ }\textbf {\bibinfo {volume} {9}},\ \bibinfo {pages} {372} (\bibinfo {year} {2014})}\BibitemShut {NoStop}%
\bibitem [{\citenamefont {Liu}\ \emph {et~al.}(2014)\citenamefont {Liu}, \citenamefont {Neal}, \citenamefont {Zhu}, \citenamefont {Luo}, \citenamefont {Xu}, \citenamefont {Tománek},\ and\ \citenamefont {Ye}}]{Liu2014}%
  \BibitemOpen
  \bibfield  {author} {\bibinfo {author} {\bibfnamefont {H.}~\bibnamefont {Liu}}, \bibinfo {author} {\bibfnamefont {A.~T.}\ \bibnamefont {Neal}}, \bibinfo {author} {\bibfnamefont {Z.}~\bibnamefont {Zhu}}, \bibinfo {author} {\bibfnamefont {Z.}~\bibnamefont {Luo}}, \bibinfo {author} {\bibfnamefont {X.}~\bibnamefont {Xu}}, \bibinfo {author} {\bibfnamefont {D.}~\bibnamefont {Tománek}},\ and\ \bibinfo {author} {\bibfnamefont {P.~D.}\ \bibnamefont {Ye}},\ }\bibfield  {title} {\bibinfo {title} {Phosphorene: An unexplored 2{D} semiconductor with a high hole mobility},\ }\href {https://doi.org/10.1021/nn501226z} {\bibfield  {journal} {\bibinfo  {journal} {ACS Nano}\ }\textbf {\bibinfo {volume} {8}},\ \bibinfo {pages} {4033} (\bibinfo {year} {2014})}\BibitemShut {NoStop}%
\bibitem [{\citenamefont {Lu}\ \emph {et~al.}(2016)\citenamefont {Lu}, \citenamefont {Yang}, \citenamefont {Carvalho}, \citenamefont {Liu}, \citenamefont {Lu},\ and\ \citenamefont {Sow}}]{Lu2016}%
  \BibitemOpen
  \bibfield  {author} {\bibinfo {author} {\bibfnamefont {J.}~\bibnamefont {Lu}}, \bibinfo {author} {\bibfnamefont {J.}~\bibnamefont {Yang}}, \bibinfo {author} {\bibfnamefont {A.}~\bibnamefont {Carvalho}}, \bibinfo {author} {\bibfnamefont {H.}~\bibnamefont {Liu}}, \bibinfo {author} {\bibfnamefont {Y.}~\bibnamefont {Lu}},\ and\ \bibinfo {author} {\bibfnamefont {C.~H.}\ \bibnamefont {Sow}},\ }\bibfield  {title} {\bibinfo {title} {Light–matter interactions in phosphorene},\ }\href {https://doi.org/10.1021/acs.accounts.6b00266} {\bibfield  {journal} {\bibinfo  {journal} {Accounts of Chemical Research}\ }\textbf {\bibinfo {volume} {49}},\ \bibinfo {pages} {1806} (\bibinfo {year} {2016})}\BibitemShut {NoStop}%
\bibitem [{\citenamefont {Rudenko}\ and\ \citenamefont {Katsnelson}(2014)}]{Rudenko2014}%
  \BibitemOpen
  \bibfield  {author} {\bibinfo {author} {\bibfnamefont {A.~N.}\ \bibnamefont {Rudenko}}\ and\ \bibinfo {author} {\bibfnamefont {M.~I.}\ \bibnamefont {Katsnelson}},\ }\bibfield  {title} {\bibinfo {title} {Quasiparticle band structure and tight-binding model for single- and bilayer black phosphorus},\ }\href {https://doi.org/10.1103/PhysRevB.89.201408} {\bibfield  {journal} {\bibinfo  {journal} {Physical Review B}\ }\textbf {\bibinfo {volume} {89}},\ \bibinfo {pages} {201408} (\bibinfo {year} {2014})}\BibitemShut {NoStop}%
\bibitem [{\citenamefont {Menezes}\ and\ \citenamefont {Capaz}(2018)}]{Menezes2018}%
  \BibitemOpen
  \bibfield  {author} {\bibinfo {author} {\bibfnamefont {M.~G.}\ \bibnamefont {Menezes}}\ and\ \bibinfo {author} {\bibfnamefont {R.~B.}\ \bibnamefont {Capaz}},\ }\bibfield  {title} {\bibinfo {title} {Tight binding parametrization of few-layer black phosphorus from first-principles calculations},\ }\href {https://doi.org/https://doi.org/10.1016/j.commatsci.2017.11.039} {\bibfield  {journal} {\bibinfo  {journal} {Computational Materials Science}\ }\textbf {\bibinfo {volume} {143}},\ \bibinfo {pages} {411} (\bibinfo {year} {2018})}\BibitemShut {NoStop}%
\bibitem [{\citenamefont {Rodin}\ \emph {et~al.}(2014)\citenamefont {Rodin}, \citenamefont {Carvalho},\ and\ \citenamefont {Castro~Neto}}]{Rodin2014}%
  \BibitemOpen
  \bibfield  {author} {\bibinfo {author} {\bibfnamefont {A.~S.}\ \bibnamefont {Rodin}}, \bibinfo {author} {\bibfnamefont {A.}~\bibnamefont {Carvalho}},\ and\ \bibinfo {author} {\bibfnamefont {A.~H.}\ \bibnamefont {Castro~Neto}},\ }\bibfield  {title} {\bibinfo {title} {Strain-induced gap modification in black phosphorus},\ }\href {https://doi.org/10.1103/PhysRevLett.112.176801} {\bibfield  {journal} {\bibinfo  {journal} {Physical Review Letters}\ }\textbf {\bibinfo {volume} {112}},\ \bibinfo {pages} {176801} (\bibinfo {year} {2014})}\BibitemShut {NoStop}%
\bibitem [{\citenamefont {Taghizadeh~Sisakht}\ \emph {et~al.}(2016)\citenamefont {Taghizadeh~Sisakht}, \citenamefont {Fazileh}, \citenamefont {Zare}, \citenamefont {Zarenia},\ and\ \citenamefont {Peeters}}]{Taghizadeh2016}%
  \BibitemOpen
  \bibfield  {author} {\bibinfo {author} {\bibfnamefont {E.}~\bibnamefont {Taghizadeh~Sisakht}}, \bibinfo {author} {\bibfnamefont {F.}~\bibnamefont {Fazileh}}, \bibinfo {author} {\bibfnamefont {M.~H.}\ \bibnamefont {Zare}}, \bibinfo {author} {\bibfnamefont {M.}~\bibnamefont {Zarenia}},\ and\ \bibinfo {author} {\bibfnamefont {F.~M.}\ \bibnamefont {Peeters}},\ }\bibfield  {title} {\bibinfo {title} {Strain-induced topological phase transition in phosphorene and in phosphorene nanoribbons},\ }\href {https://doi.org/10.1103/PhysRevB.94.085417} {\bibfield  {journal} {\bibinfo  {journal} {Physical Review B}\ }\textbf {\bibinfo {volume} {94}},\ \bibinfo {pages} {085417} (\bibinfo {year} {2016})}\BibitemShut {NoStop}%
\bibitem [{\citenamefont {Midtvedt}\ \emph {et~al.}(2017)\citenamefont {Midtvedt}, \citenamefont {Lewenkopf},\ and\ \citenamefont {Croy}}]{Midtvedt2017}%
  \BibitemOpen
  \bibfield  {author} {\bibinfo {author} {\bibfnamefont {D.}~\bibnamefont {Midtvedt}}, \bibinfo {author} {\bibfnamefont {C.~H.}\ \bibnamefont {Lewenkopf}},\ and\ \bibinfo {author} {\bibfnamefont {A.}~\bibnamefont {Croy}},\ }\bibfield  {title} {\bibinfo {title} {Multi-scale approach for strain-engineering of phosphorene},\ }\href {https://doi.org/10.1088/1361-648X/aa66d4} {\bibfield  {journal} {\bibinfo  {journal} {Journal of Physics: Condensed Matter}\ }\textbf {\bibinfo {volume} {29}},\ \bibinfo {pages} {185702} (\bibinfo {year} {2017})}\BibitemShut {NoStop}%
\bibitem [{\citenamefont {Nemilentsau}\ \emph {et~al.}(2016)\citenamefont {Nemilentsau}, \citenamefont {Low},\ and\ \citenamefont {Hanson}}]{Nemilentsau2016}%
  \BibitemOpen
  \bibfield  {author} {\bibinfo {author} {\bibfnamefont {A.}~\bibnamefont {Nemilentsau}}, \bibinfo {author} {\bibfnamefont {T.}~\bibnamefont {Low}},\ and\ \bibinfo {author} {\bibfnamefont {G.}~\bibnamefont {Hanson}},\ }\bibfield  {title} {\bibinfo {title} {Anisotropic 2{D} materials for tunable hyperbolic plasmonics},\ }\href {https://doi.org/10.1103/PhysRevLett.116.066804} {\bibfield  {journal} {\bibinfo  {journal} {Physical Review Letters}\ }\textbf {\bibinfo {volume} {116}},\ \bibinfo {pages} {066804} (\bibinfo {year} {2016})}\BibitemShut {NoStop}%
\bibitem [{\citenamefont {van Veen}\ \emph {et~al.}(2019)\citenamefont {van Veen}, \citenamefont {Nemilentsau}, \citenamefont {Kumar}, \citenamefont {Rold\'an}, \citenamefont {Katsnelson}, \citenamefont {Low},\ and\ \citenamefont {Yuan}}]{Veen2019}%
  \BibitemOpen
  \bibfield  {author} {\bibinfo {author} {\bibfnamefont {E.}~\bibnamefont {van Veen}}, \bibinfo {author} {\bibfnamefont {A.}~\bibnamefont {Nemilentsau}}, \bibinfo {author} {\bibfnamefont {A.}~\bibnamefont {Kumar}}, \bibinfo {author} {\bibfnamefont {R.}~\bibnamefont {Rold\'an}}, \bibinfo {author} {\bibfnamefont {M.~I.}\ \bibnamefont {Katsnelson}}, \bibinfo {author} {\bibfnamefont {T.}~\bibnamefont {Low}},\ and\ \bibinfo {author} {\bibfnamefont {S.}~\bibnamefont {Yuan}},\ }\bibfield  {title} {\bibinfo {title} {Tuning two-dimensional hyperbolic plasmons in black phosphorus},\ }\href {https://doi.org/10.1103/PhysRevApplied.12.014011} {\bibfield  {journal} {\bibinfo  {journal} {Physical Review Applied}\ }\textbf {\bibinfo {volume} {12}},\ \bibinfo {pages} {014011} (\bibinfo {year} {2019})}\BibitemShut {NoStop}%
\bibitem [{\citenamefont {Sun}\ \emph {et~al.}(2017)\citenamefont {Sun}, \citenamefont {Zhang}, \citenamefont {Zhang},\ and\ \citenamefont {Ji}}]{Sun2017}%
  \BibitemOpen
  \bibfield  {author} {\bibinfo {author} {\bibfnamefont {L.}~\bibnamefont {Sun}}, \bibinfo {author} {\bibfnamefont {G.}~\bibnamefont {Zhang}}, \bibinfo {author} {\bibfnamefont {S.}~\bibnamefont {Zhang}},\ and\ \bibinfo {author} {\bibfnamefont {J.}~\bibnamefont {Ji}},\ }\bibfield  {title} {\bibinfo {title} {Enhanced spontaneous emission of quantum emitter in monolayer and double layer black phosphorus},\ }\href {https://opg.optica.org/oe/abstract.cfm?URI=oe-25-13-14270} {\bibfield  {journal} {\bibinfo  {journal} {Optics Express}\ }\textbf {\bibinfo {volume} {25}},\ \bibinfo {pages} {14270} (\bibinfo {year} {2017})}\BibitemShut {NoStop}%
\bibitem [{\citenamefont {Thiyam}\ \emph {et~al.}(2018)\citenamefont {Thiyam}, \citenamefont {Parashar}, \citenamefont {Shajesh}, \citenamefont {Malyi}, \citenamefont {Bostr\"om}, \citenamefont {Milton}, \citenamefont {Brevik},\ and\ \citenamefont {Persson}}]{Thiyam2018}%
  \BibitemOpen
  \bibfield  {author} {\bibinfo {author} {\bibfnamefont {P.}~\bibnamefont {Thiyam}}, \bibinfo {author} {\bibfnamefont {P.}~\bibnamefont {Parashar}}, \bibinfo {author} {\bibfnamefont {K.~V.}\ \bibnamefont {Shajesh}}, \bibinfo {author} {\bibfnamefont {O.~I.}\ \bibnamefont {Malyi}}, \bibinfo {author} {\bibfnamefont {M.}~\bibnamefont {Bostr\"om}}, \bibinfo {author} {\bibfnamefont {K.~A.}\ \bibnamefont {Milton}}, \bibinfo {author} {\bibfnamefont {I.}~\bibnamefont {Brevik}},\ and\ \bibinfo {author} {\bibfnamefont {C.}~\bibnamefont {Persson}},\ }\bibfield  {title} {\bibinfo {title} {Distance-dependent sign reversal in the {C}asimir-{L}ifshitz torque},\ }\href {https://doi.org/10.1103/PhysRevLett.120.131601} {\bibfield  {journal} {\bibinfo  {journal} {Physical Review Letters}\ }\textbf {\bibinfo {volume} {120}},\ \bibinfo {pages} {131601} (\bibinfo {year} {2018})}\BibitemShut {NoStop}%
\bibitem [{\citenamefont {Mu}\ \emph {et~al.}(2021)\citenamefont {Mu}, \citenamefont {Wang}, \citenamefont {Zhang}, \citenamefont {Liu}, \citenamefont {Yu},\ and\ \citenamefont {Liao}}]{Mu2021}%
  \BibitemOpen
  \bibfield  {author} {\bibinfo {author} {\bibfnamefont {H.}~\bibnamefont {Mu}}, \bibinfo {author} {\bibfnamefont {T.}~\bibnamefont {Wang}}, \bibinfo {author} {\bibfnamefont {D.}~\bibnamefont {Zhang}}, \bibinfo {author} {\bibfnamefont {W.}~\bibnamefont {Liu}}, \bibinfo {author} {\bibfnamefont {T.}~\bibnamefont {Yu}},\ and\ \bibinfo {author} {\bibfnamefont {Q.}~\bibnamefont {Liao}},\ }\bibfield  {title} {\bibinfo {title} {Mechanical modulation of spontaneous emission of nearby nanostructured black phosphorus},\ }\href {https://doi.org/10.1364/OE.414380} {\bibfield  {journal} {\bibinfo  {journal} {Optics Express}\ }\textbf {\bibinfo {volume} {29}},\ \bibinfo {pages} {1037} (\bibinfo {year} {2021})}\BibitemShut {NoStop}%
\bibitem [{\citenamefont {Sikder}\ \emph {et~al.}(2022)\citenamefont {Sikder}, \citenamefont {Nayem},\ and\ \citenamefont {Uddin}}]{Sikder2022}%
  \BibitemOpen
  \bibfield  {author} {\bibinfo {author} {\bibfnamefont {B.}~\bibnamefont {Sikder}}, \bibinfo {author} {\bibfnamefont {S.~H.}\ \bibnamefont {Nayem}},\ and\ \bibinfo {author} {\bibfnamefont {S.~Z.}\ \bibnamefont {Uddin}},\ }\bibfield  {title} {\bibinfo {title} {Deep ultraviolet spontaneous emission enhanced by layer dependent black phosphorus plasmonics},\ }\href {https://doi.org/10.1364/OE.478735} {\bibfield  {journal} {\bibinfo  {journal} {Optics Express}\ }\textbf {\bibinfo {volume} {30}},\ \bibinfo {pages} {47152} (\bibinfo {year} {2022})}\BibitemShut {NoStop}%
\bibitem [{\citenamefont {Tao}\ \emph {et~al.}(2024)\citenamefont {Tao}, \citenamefont {Lavor}, \citenamefont {Dong}, \citenamefont {Chaves}, \citenamefont {Neilson},\ and\ \citenamefont {Milošević}}]{Tao2024}%
  \BibitemOpen
  \bibfield  {author} {\bibinfo {author} {\bibfnamefont {Z.}~\bibnamefont {Tao}}, \bibinfo {author} {\bibfnamefont {I.~R.}\ \bibnamefont {Lavor}}, \bibinfo {author} {\bibfnamefont {H.}~\bibnamefont {Dong}}, \bibinfo {author} {\bibfnamefont {A.}~\bibnamefont {Chaves}}, \bibinfo {author} {\bibfnamefont {D.}~\bibnamefont {Neilson}},\ and\ \bibinfo {author} {\bibfnamefont {M.~V.}\ \bibnamefont {Milošević}},\ }\bibfield  {title} {\bibinfo {title} {Chiral propagation of plasmon polaritons due to competing anisotropies in a twisted photonic heterostructure},\ }\href {https://pubs.acs.org/doi/abs/10.1021/acs.nanolett.4c04502} {\bibfield  {journal} {\bibinfo  {journal} {Nano Letters}\ }\textbf {\bibinfo {volume} {24}},\ \bibinfo {pages} {15745} (\bibinfo {year} {2024})}\BibitemShut {NoStop}%
\bibitem [{\citenamefont {Born}\ and\ \citenamefont {Wolf}(2019)}]{Born2019}%
  \BibitemOpen
  \bibfield  {author} {\bibinfo {author} {\bibfnamefont {M.}~\bibnamefont {Born}}\ and\ \bibinfo {author} {\bibfnamefont {E.}~\bibnamefont {Wolf}},\ }\href@noop {} {\emph {\bibinfo {title} {Principles of Optics: 60th Anniversary Edition}}},\ \bibinfo {edition} {7th}\ ed.\ (\bibinfo  {publisher} {Cambridge University Press},\ \bibinfo {year} {2019})\BibitemShut {NoStop}%
\bibitem [{\citenamefont {Amorim}\ \emph {et~al.}(2017)\citenamefont {Amorim}, \citenamefont {Gonçalves}, \citenamefont {Vasilevskiy},\ and\ \citenamefont {Peres}}]{Amorim2017}%
  \BibitemOpen
  \bibfield  {author} {\bibinfo {author} {\bibfnamefont {B.}~\bibnamefont {Amorim}}, \bibinfo {author} {\bibfnamefont {P.~A.~D.}\ \bibnamefont {Gonçalves}}, \bibinfo {author} {\bibfnamefont {M.~I.}\ \bibnamefont {Vasilevskiy}},\ and\ \bibinfo {author} {\bibfnamefont {N.~M.~R.}\ \bibnamefont {Peres}},\ }\bibfield  {title} {\bibinfo {title} {Impact of graphene on the polarizability of a neighbour nanoparticle: A dyadic {G}reen’s function study},\ }\href {https://www.mdpi.com/2076-3417/7/11/1158} {\bibfield  {journal} {\bibinfo  {journal} {Applied Sciences}\ }\textbf {\bibinfo {volume} {7}},\ \bibinfo {pages} {1158} (\bibinfo {year} {2017})}\BibitemShut {NoStop}%
\bibitem [{\citenamefont {Karanikolas}\ \emph {et~al.}(2016)\citenamefont {Karanikolas}, \citenamefont {Marocico},\ and\ \citenamefont {Bradley}}]{Karanikolas2016}%
  \BibitemOpen
  \bibfield  {author} {\bibinfo {author} {\bibfnamefont {V.~D.}\ \bibnamefont {Karanikolas}}, \bibinfo {author} {\bibfnamefont {C.~A.}\ \bibnamefont {Marocico}},\ and\ \bibinfo {author} {\bibfnamefont {A.~L.}\ \bibnamefont {Bradley}},\ }\bibfield  {title} {\bibinfo {title} {Tunable and long-range energy transfer efficiency through a graphene nanodisk},\ }\href {https://doi.org/10.1103/PhysRevB.93.035426} {\bibfield  {journal} {\bibinfo  {journal} {Physical Review B}\ }\textbf {\bibinfo {volume} {93}},\ \bibinfo {pages} {035426} (\bibinfo {year} {2016})}\BibitemShut {NoStop}%
\bibitem [{\citenamefont {Ding}\ \emph {et~al.}(2017)\citenamefont {Ding}, \citenamefont {Hsu},\ and\ \citenamefont {Schatz}}]{Ding2017}%
  \BibitemOpen
  \bibfield  {author} {\bibinfo {author} {\bibfnamefont {W.}~\bibnamefont {Ding}}, \bibinfo {author} {\bibfnamefont {L.-Y.}\ \bibnamefont {Hsu}},\ and\ \bibinfo {author} {\bibfnamefont {G.~C.}\ \bibnamefont {Schatz}},\ }\bibfield  {title} {\bibinfo {title} {Plasmon-coupled resonance energy transfer: A real-time electrodynamics approach},\ }\href {http://dx.doi.org/10.1063/1.4975815} {\bibfield  {journal} {\bibinfo  {journal} {The Journal of Chemical Physics}\ }\textbf {\bibinfo {volume} {146}},\ \bibinfo {pages} {064109} (\bibinfo {year} {2017})}\BibitemShut {NoStop}%
\bibitem [{\citenamefont {Wu}\ \emph {et~al.}(2018)\citenamefont {Wu}, \citenamefont {Lin}, \citenamefont {Sheu},\ and\ \citenamefont {Hsu}}]{Wu2018}%
  \BibitemOpen
  \bibfield  {author} {\bibinfo {author} {\bibfnamefont {J.-S.}\ \bibnamefont {Wu}}, \bibinfo {author} {\bibfnamefont {Y.-C.}\ \bibnamefont {Lin}}, \bibinfo {author} {\bibfnamefont {Y.-L.}\ \bibnamefont {Sheu}},\ and\ \bibinfo {author} {\bibfnamefont {L.-Y.}\ \bibnamefont {Hsu}},\ }\bibfield  {title} {\bibinfo {title} {Characteristic distance of resonance energy transfer coupled with surface plasmon polaritons},\ }\href {https://doi.org/10.1021/acs.jpclett.8b03429} {\bibfield  {journal} {\bibinfo  {journal} {The Journal of Physical Chemistry Letters}\ }\textbf {\bibinfo {volume} {9}},\ \bibinfo {pages} {7032} (\bibinfo {year} {2018})}\BibitemShut {NoStop}%
\bibitem [{\citenamefont {Rodriguez-Lopez}\ \emph {et~al.}(2024)\citenamefont {Rodriguez-Lopez}, \citenamefont {Le}, \citenamefont {Bondarev}, \citenamefont {Antezza},\ and\ \citenamefont {Woods}}]{Rodriguez-Lopes2024}%
  \BibitemOpen
  \bibfield  {author} {\bibinfo {author} {\bibfnamefont {P.}~\bibnamefont {Rodriguez-Lopez}}, \bibinfo {author} {\bibfnamefont {D.-N.}\ \bibnamefont {Le}}, \bibinfo {author} {\bibfnamefont {I.~V.}\ \bibnamefont {Bondarev}}, \bibinfo {author} {\bibfnamefont {M.}~\bibnamefont {Antezza}},\ and\ \bibinfo {author} {\bibfnamefont {L.~M.}\ \bibnamefont {Woods}},\ }\bibfield  {title} {\bibinfo {title} {Giant anisotropy and {C}asimir phenomena: The case of carbon nanotube metasurfaces},\ }\href {https://doi.org/10.1103/PhysRevB.109.035422} {\bibfield  {journal} {\bibinfo  {journal} {Physical Review B}\ }\textbf {\bibinfo {volume} {109}},\ \bibinfo {pages} {035422} (\bibinfo {year} {2024})}\BibitemShut {NoStop}%
\bibitem [{\citenamefont {Li}\ \emph {et~al.}(2021)\citenamefont {Li}, \citenamefont {Liu}, \citenamefont {Ke}, \citenamefont {Tang}, \citenamefont {Liu}, \citenamefont {Huang}, \citenamefont {Wu}, \citenamefont {Wu},\ and\ \citenamefont {Kang}}]{Li2021}%
  \BibitemOpen
  \bibfield  {author} {\bibinfo {author} {\bibfnamefont {X.}~\bibnamefont {Li}}, \bibinfo {author} {\bibfnamefont {H.}~\bibnamefont {Liu}}, \bibinfo {author} {\bibfnamefont {C.}~\bibnamefont {Ke}}, \bibinfo {author} {\bibfnamefont {W.}~\bibnamefont {Tang}}, \bibinfo {author} {\bibfnamefont {M.}~\bibnamefont {Liu}}, \bibinfo {author} {\bibfnamefont {F.}~\bibnamefont {Huang}}, \bibinfo {author} {\bibfnamefont {Y.}~\bibnamefont {Wu}}, \bibinfo {author} {\bibfnamefont {Z.}~\bibnamefont {Wu}},\ and\ \bibinfo {author} {\bibfnamefont {J.}~\bibnamefont {Kang}},\ }\bibfield  {title} {\bibinfo {title} {Review of anisotropic 2{D} materials: Controlled growth, optical anisotropy modulation, and photonic applications},\ }\href {https://doi.org/https://doi.org/10.1002/lpor.202100322} {\bibfield  {journal} {\bibinfo  {journal} {Laser \& Photonics Reviews}\ }\textbf {\bibinfo {volume} {15}},\ \bibinfo {pages} {2100322} (\bibinfo {year} {2021})}\BibitemShut {NoStop}%
\bibitem [{\citenamefont {Wald}(2022)}]{Wald}%
  \BibitemOpen
  \bibfield  {author} {\bibinfo {author} {\bibfnamefont {R.}~\bibnamefont {Wald}},\ }\href {https://books.google.com.br/books?id=y7-PzgEACAAJ} {\emph {\bibinfo {title} {Advanced Classical Electromagnetism}}}\ (\bibinfo  {publisher} {Princeton University Press},\ \bibinfo {year} {2022})\BibitemShut {NoStop}%
\bibitem [{\citenamefont {Palik}(1998)}]{palik1998}%
  \BibitemOpen
  \bibfield  {author} {\bibinfo {author} {\bibfnamefont {E.~D.}\ \bibnamefont {Palik}},\ }\href@noop {} {\emph {\bibinfo {title} {Handbook of optical constants of solids}}},\ Vol.~\bibinfo {volume} {3}\ (\bibinfo  {publisher} {Academic press},\ \bibinfo {year} {1998})\BibitemShut {NoStop}%
\end{thebibliography}

\end{document}